\documentclass[a4paper,titlepage,12pt]{article}

\usepackage{arxiv}

\usepackage[utf8]{inputenc} 					
\usepackage[english]{babel}
\usepackage[T1]{fontenc}                    
\usepackage{xcolor}
\definecolor{darkred}{rgb}{0.55, 0.0, 0.0}
\definecolor{darkcerulean}{rgb}{0.03, 0.27, 0.49}
\usepackage[colorlinks]{hyperref}          
\hypersetup{urlcolor={darkred}, citecolor={darkcerulean}}
\usepackage{url}                   					
\usepackage{booktabs}         					
\usepackage{titlesec}
\titlelabel{\thetitle.\quad}

\usepackage{amssymb,amsmath,amsfonts} 
\usepackage{calrsfs}
\usepackage{siunitx}     

\usepackage{nicefrac}       
\usepackage{microtype}      
\usepackage{lipsum}		

\usepackage[numbers]{natbib}

\usepackage[inline]{enumitem}

\usepackage{graphicx}
\usepackage{caption}
\usepackage{subcaption}

\usepackage{pgfplots}
\usepackage{pgfplotstable}
\pgfplotsset{compat=newest}
\usepackage{pgfplots}
\usetikzlibrary{fadings}
\usepgfplotslibrary{fillbetween}
\PassOptionsToPackage{gray}{xcolor}
\definecolor{orange}{gray}{0.4}

\usepackage{color}
\definecolor{red}{rgb}{0.5, 0.0, 0.13}

\pgfdeclarehorizontalshading{FOne}{100bp}
{color(0bp)=(pgftransparent!100); color(25bp)=(pgftransparent!100);
 color(75bp)=(pgftransparent!0); color(100bp)=(pgftransparent!0)}

\pgfdeclarefading{myfading}{\pgfuseshading{FOne}}

\usepgfplotslibrary{polar}
\usepgflibrary{shapes.geometric}
\usetikzlibrary{calc}

\usepackage{tikz}
\usetikzlibrary{positioning}
\usepackage{tikzscale}
\usetikzlibrary{calc}

\title{Simplex space-time meshes in two-phase flow simulations}

\date{} 					

\author{
Violeta~Karyofylli\thanks{Corresponding author\newline \textit{Email addresses:} \href{mailto:karyofylli@cats.rwth-aachen.de}{karyofylli@cats.rwth-aachen.de} (Violeta Karyofylli), \href{mailto:frings@cats.rwth-aachen.de}{frings@cats.rwth-aachen.de} (Markus Frings), \href{mailto:elgeti@cats.rwth-aachen.de}{elgeti@cats.rwth-aachen.de} (Stefanie Elgeti), \href{mailto:behr@cats.rwth-aachen.de}{behr@cats.rwth-aachen.de} (Marek Behr) \newline
\newline \textit{NOTICE:} This is the author’s version of a work that was accepted for publication in \textit{International Journal for Numerical Methods in Fluids}. Changes resulting from the publishing process, such as editing, corrections, structural formatting, and other quality control mechanisms may not be reflected in this document. Changes may have been made to this work since it was submitted for publication.}\ , \ Markus~Frings, \ Stefanie~Elgeti, \ Marek~Behr}

\begin{document}
\maketitle

\begin{abstract}
In this paper, we present the numerical solution of two-phase flow problems of engineering significance with a space-time finite element method that allows for local temporal refinement. Our basis is the method presented in \citep{behr2008simplex}, which allows for arbitrary temporal refinement in preselected regions of the mesh. It has been extended to adaptive temporal refinement that is governed by a quantity that is part of the solution process, namely, the interface position in two-phase flow. Due to local effects such as surface tension, jumps in material properties, etc., the interface can, in general, be considered a region that requires high flexibility and high resolution, both in space and in time. The new method, which leads to tetrahedral (for 2D problems) and pentatope (for 3D problems) meshes, offers an efficient yet accurate approach to the underlying two-phase flow problems.   
\end{abstract}

\keywords{discontinuous-in-time Galerkin \and space-time finite elements \and space-time \and simplex \and evolving front \and two-phase flow \and level-set}

\section{Introduction}
\label{Introduction}
The spatial discretization of a time-dependent problem is often performed by means of the Galerkin or Petrov-Galerkin finite element (FE) method. However, the time discretization is typically being based on an explicit or implicit finite difference (FD) temporal discretization, such as \(\theta\)-family schemes or Runge-Kutta-family of methods. In recent years, the space-time finite element method has been steadily applied to more and more problems, e.g., advective-diffusive systems \cite{hughes1989new,hughes1987new}, elastodynamics \cite{hughes1988space}, Navier-Stokes equations \cite{shakib1989finite,hansbo1990velocity} and Navier-Stokes equations with deforming domains \cite{tezduyar1992newA,tezduyar1992newB,hansbo1992characteristic}.

The space-time approach utilizes subsets of the temporal domain called space-time slabs, which are more or less similar to time steps of the standard semi-discrete approach. In most space-time implementations so far, the meshes for the space-time slabs are simply extruded in the temporal direction from a spatial mesh, resulting in semi-unstructured domains, which can be either unstructured or structured in space but structured in time. That means that space-time method has inherent flexibility to admit completely unstructured meshes with varying levels of refinement only in spatial dimensions, but does not allow different temporal refinement in different parts of the domain.

Nowadays, the extraordinary flexibility of the space-time FE is being exploited, when dealing with varying resolution of complicated domains not only in space dimensions but also in the time dimension. That leads to the use of fully unstructured meshes in both space and time. The construction of simplicial meshes suitable for space-time discontinuous Galerkin finite-element methods was introduced in \cite{erickson2005building} and relied on the ``Tent Pitcher'' algorithm of \cite{ungor2000tent}. In \cite{behr2008simplex}, the generation of simplex space-time meshes was demonstrated and tested in the context of the advection-diffusion equation. \citet{wang2013discontinuous} showed a fully unstructured space-time mesh, which can cope with any type of domain deformations, even with topological changes. They also used local mesh operations in order to avoid remeshing. An algorithm for arbitrary finite element discretizations of the space-time cylinder was presented by \citet{neumuller2011refinement} as well. This method does not depend on the time-slabs, resulting in adaptive meshes, movable in time. In \cite{neumuller2011refinement}, the decomposition of a pentatope into smaller ones is also proposed. This decomposition relies on the Freudenthal algorithm \cite{freudenthal1942simplizialzerlegungen}. In \cite{lehrenfeld2015nitsche}, a new procedure for the subdivision of four-dimensional prisms intersected by a moving front into simplices was illustrated. Such a subdivision method is important for a better resolution of the space-time interface.

In the present paper, we show a fully-unstructured space-time discretization of an interface-capturing finite element method, designed for two-phase incompressible flows including surface tension effects. We use \(P1P1\) finite elements with least-squares stabilization. This approach is based on the discontinuous-Galerkin method in time (space-time elements), details of which can be found in \cite{hughes1989new,hughes1987new}. The variational formulation of the problem is written over the associated space-time domain. The interface is approximated by the level-set method. Level-set method describes implicitly the interface, meaning that the formulation is able to cope with extreme topological changes of the evolving front between the two phases. The benchmark cases reveal that the simulation results obtained with the fully-unstructured space-time discretization are equivalent to those obtained with the standard discretization, but offer a potential reduction in the number of degrees of freedom. 

The structure of this paper is the following: In Section \(\SI{2}{}\) and \(\SI{3}{}\), the governing equations and their discretization are described, respectively. Section \(\SI{4}{}\) deals with the generation of simplex-type space-time meshes. In Section \(\SI{5}{}\), the numerical results are presented. The concluding remarks are drawn in Section \(\SI{6}{}\).

\section{Governing equations}
\label{GoverningEquations}

We focus on the incompressible two-phase flow which is governed by the transient, isothermal and incompressible Navier-Stokes equations. A computational domain \(\Omega\) is considered, which is a subset of \(\mathbb{R}^{n_{sd}}\); \(n_{sd}\) is the number of space dimensions. This domain encloses two immiscible Newtonian phases \(\Omega_1(t)\) and \(\Omega_2(t)\), where \(\Omega_1(t)\cup\Omega_2(t) = \Omega\). The boundary of the domain is denoted by \(\Gamma = \partial{\Omega}\), whereas \(\Gamma_{int}(t)\) stands for the interface between the two fluids \(\partial{\Omega_1(t)}\cap\partial{\Omega_2(t)}\). Note that the spatial subdomains of the two phases and their interface are time-dependent.

At each instant \(t \in [0,T]\), the velocity, \(\bold{u}(\bold{x}, t )\), and the pressure, \(p(\bold{x}, t )\), in each phase are governed by the following
equations:
\begin{equation} \label{Navier_Stokes}
\rho_i\left(\frac{\partial \bold{u}}{\partial t}+\bold{u}\cdot\nabla\bold{u}-\bold{f}\right) -\nabla\cdot\boldsymbol{\sigma}_i= 0 \quad  \mathrm{in} \quad  \Omega_i(t),
\end{equation}
\begin{equation} \label{Continuity_eq}
\nabla\cdot\bold{u}= 0 \quad  \mathrm{in} \quad  \Omega_i(t),
\end{equation}
for \(i =1, 2\) number of phases. Here, \(\rho_i\) is the density of the corresponding fluid. The stress tensor \(\sigma_i\) and the rate of strain tensor are defined as
\begin{equation} \label{Stress_tensor}
\boldsymbol{\sigma}_i(\bold{u},p) = -p\bold{I}+2\mu_i\boldsymbol{\varepsilon}(\bold{u}) \quad  \mathrm{in} \quad  \Omega_i(t), \quad  i =1, 2 ,
\end{equation}
\begin{equation} \label{Strain_tensor}
\boldsymbol{\varepsilon}(\bold{u})= \frac{1}{2}(\nabla\bold{u}+(\nabla\bold{u})^T ) ,
\end{equation}
where \(\mu_i\) is the dynamic viscosity of each fluid.

At the interface between the two phases, we impose a boundary condition which is based on the Laplace-Young equation:
\begin{equation} \label{Surface_tension}
\bold{n}\cdot[\boldsymbol{\sigma}]_{\Gamma_{int}(t)}=\gamma\kappa\bold{n} \quad  \mathrm{on} \quad  \Gamma_{int}(t), \quad  \forall t \in [0,T] ,
\end{equation}
where \(\bold{n}\) is the outward unit normal vector on \(\Gamma_{int}(t)\), \(\gamma\) the surface tension coefficient and \(\kappa\) the local curvature of \(\Gamma_{int}(t)\). Furthermore, the velocities are assumed to be continuous across the interface:
\begin{equation} \label{Surface_tension}
[\bold{u}]_{\Gamma_{int}(t)}=\bold{0} \quad  \mathrm{on} \quad  \Gamma_{int}(t), \quad  \forall t \in [0,T] .
\end{equation}

In order to elucidate the evolution of the interface between the two phases, the level-set transport equation \cite{osher1988fronts}:
\begin{equation} \label{Level-Set}
\frac{\partial\phi}{\partial t}+\bold{u}\cdot\nabla\phi = 0 \quad  \mathrm{in} \quad  \Omega_i(t), \quad  \forall t \in [0,T] ,
\end{equation}
is solved, where \(\phi\) is a signed-distance function and \(\bold{u}\) is the velocity field obtained from the Navier-Stokes equations (\ref{Navier_Stokes}) -- (\ref{Continuity_eq}).
\section{Solution technique}
\label{SolutionTechnique}
For the discretization of equations (\ref{Navier_Stokes}), (\ref{Continuity_eq}) and (\ref{Level-Set}), we use a first-order interpolation of all degrees of freedom in combination with a Galerkin/Least-Squares (GLS) stabilization in space and DG in time \cite{donea2003finite}. For each space-time slab \(Q_n\), the following finite element interpolation and weighting function spaces are defined for the velocity, the pressure and the level-set function:
\begin{equation} \label{Interpolation_fs_vel}
(\boldsymbol{\mathcal{S}}_\bold{u}^h)_n = \{\bold{u}^h | \bold{u}^h \in [H^{1h}(Q_n)]^{n_{sd}}, \bold{u}^h \dot{=} \bold{g}^h \quad \mathrm{on} \quad (P_n)_\bold{g}\} ,
\end{equation}
\begin{equation} \label{Weighting_fs_vel}
(\mathcal{V}_\bold{u}^h)_n = \{\bold{w}^h | \bold{w}^h \in [H^{1h}(Q_n)]^{n_{sd}}, \bold{w}^h \dot{=} \bold{0} \quad \mathrm{on} \quad (P_n)_\bold{g}\} ,
\end{equation}
\begin{equation} \label{Interpolation_Weighting_fs_pres}
(\mathcal{S}_p^h)_n = (\mathcal{V}_p^h)_n = \{p^h | p^h \in H^{1h}(Q_n)\} ,
\end{equation}
\begin{equation} \label{Interpolation_fs_fi}
(\boldsymbol{\mathcal{S}}_\phi^h)_n = \{\phi^h | \phi^h \in (H^{1h})^{n_{sd}}, \phi^h \dot{=} \hat{\phi}^h \quad \mathrm{on} \quad (P_n)_{\phi}\} ,
\end{equation}
\begin{equation} \label{Weighting_fs_fi}
(\mathcal{V}_\phi^h)_n = \{v^h | v^h \in (H^{1h})^{n_{sd}}, v^h \dot{=} 0 \quad \mathrm{on} \quad (P_n)_{\phi}\} .
\end{equation}
Here, \(P_n\) denotes the space-time extruded boundary, whereas \(\bold{g}\) and \(\hat{\phi}\) the prescribed values for the velocity and the level-set function, respectively, on the appropriate subset of \(P_n\).

The stabilized space-time formulation for the Navier-Stokes equations (\ref{Navier_Stokes}) and (\ref{Continuity_eq}) can then be written as follows: Given \((\bold{u}^h)_n^-\), find \(\bold{u}^h \in (\mathcal{S}_\bold{u}^h)_n\) and \(p^h \in (\mathcal{S}_p^h)_n\) such that \(\forall \bold{w}^h \in (\mathcal{V}_\bold{u}^h)_n, \forall q^h \in (\mathcal{V}_p^h)_n\):
\begin{equation} \label{Variational_Navier-Stokes}
\begin{aligned}
\int_{Q_n}&\bold{w}^h\cdot\rho_i\left(\frac{\partial \bold{u}^h}{\partial t}+\bold{u}^h\cdot\nabla\bold{u}^h-\bold{f}\right) \ dQ
+ \int_{Q_n}\boldsymbol{\varepsilon}(\bold{w}^h):\boldsymbol{\sigma}_i(\bold{u}^h,p^h) \ dQ \\
&+ \int_{Q_n}q^h\nabla\cdot\bold{u}^h\ dQ
+ \int_{\Omega_n}(\bold{w}^h)^+_n\cdot\rho_i((\bold{u}^h)^+_n-(\bold{u}^h)^-_n) \ d\Omega \\
&+ \sum_{e=1}^{(n_{el})_n} \int_{Q_n^e}\tau_{MOM}\frac{1}{\rho_i} \left[\rho_i \left(\frac{\partial \bold{w}^h}{\partial t}+\bold{u}^h\cdot\nabla \bold{w}^h \right)-\nabla\cdot\boldsymbol{\sigma}_i(\bold{w}^h, q^h) \right] \\
&\cdot \left[\rho_i \left(\frac{\partial \bold{u}^h}{\partial t}+\bold{u}^h\cdot\nabla \bold{u}^h -\bold{f}\right)-\nabla\cdot\boldsymbol{\sigma}_i(\bold{u}^h, p^h) \right] \ dQ \\
&+ \sum_{e=1}^{(n_{el})_n}\int_{Q_n^e}\tau_{CONT}\nabla\cdot\bold{w}^h\rho_i\nabla\cdot\bold{u}^h\ dQ  = \int_{(P_{int})_{n}}\bold{w}^h\cdot\gamma\kappa\bold{n} \ dP .
\end{aligned}
\end{equation}

The stabilized space-time formulation for the level-set transport equation (\ref{Level-Set}) can then be written as follows: Given \((\phi^h)_n^-\), find \(\phi^h \in (\mathcal{S}_\phi^h)_n\) such that \(\forall v^h \in (\mathcal{V}_\phi^h)_n\):
\begin{equation} \label{Variational_Level-Set}
\begin{aligned}
\int_{Q_n}&v^h(\frac{\partial\phi^h}{\partial t} +\bold{u}^h\cdot\nabla\phi^h) \ dQ + \int_{\Omega_n}(v^h)^+_n((\phi^h)^+_n-(\phi^h)^-_n) \ d\Omega \\
&+ \sum_{e=1}^{n_{el}} \int_{Q_n^e}(\frac{\partial v^h}{\partial t}+\bold{u}^h\cdot\nabla v^h)\tau_{LEV}(\frac{\partial\phi^h}{\partial t}+\bold{u}^h\cdot\nabla\phi^h) \ dQ = 0 .
\end{aligned}
\end{equation}
The notation \((\bullet^h)^+_n\) and \((\bullet^h)^-_n\) in the weak formulations (\ref{Variational_Navier-Stokes}) and (\ref{Variational_Level-Set}) denote the upper and lower values, respectively, of the discontinuous variable at the lower surface \(\Omega\) of the space-time slab \(Q_n\). The problem is solved sequentially for each space-time slab, starting with \((\bullet^h)^+_n=\bullet_0\). Details on the parameters \(\tau_{MOM}\) and \(\tau_{CONT}\) in (\ref{Variational_Navier-Stokes}) can be found in \cite{behr1994finite}, whereas details on the parameter \(\tau_{LEV}\) in (\ref{Variational_Level-Set}) are given by \citet{sauerland2013stable}.

The Laplace-Beltrami technique, as proposed in \cite{hysing2006new}, is employed to reformulate the surface tension term in the weak formulation (\ref{Variational_Navier-Stokes}) and results in:
\begin{equation} \label{Laplace_Beltrami}
\begin{aligned}
\int_{(P_{int})_{n}}\bold{w}^h\cdot\gamma\kappa\bold{n} \ dP &= \int_{(P_{int})_{n}}\bold{w}^h\cdot\gamma\underline\Delta\bold{id}_{(P_{int})_{n}} \ dP \\
& = - \int_{(P_{int})_{n}}\gamma\underline\nabla\bold{id}_{(P_{int})_{n}}:\underline\nabla\bold{w}^h \ dP ,
\end{aligned}
\end{equation}
where \(\underline\Delta\) is the Laplace-Beltrami operator, \(\underline\nabla\) is tangential gradient and \(\bold{id}\) is the identity mapping on the space-time evolving interface \((P_{int})_{n}\).
\section{Simplex space-time meshes}
\label{SimplexSpace-timeMeshes}
The key to space-time simulations with varying degrees of temporal refinement is the generation of simplex-based space-time meshes. A straightforward and robust algorithm that produces this type of meshes was already described by \citet{behr2008simplex}. Here, we use an updated version of this algorithm that allows arbitrary temporal refinement in selected portions of space-time slabs based on the level-set field.

\begin{figure}[!htb]
    \centering
    \begin{subfigure}[t]{0.45\textwidth}
        \centering
        \includegraphics[width=0.485\textwidth]{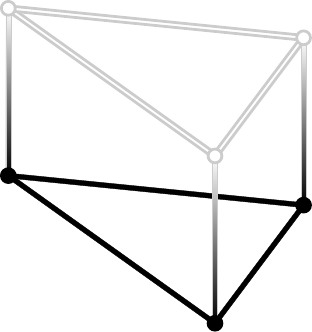}
        \caption{3d6n}
        \label{fig:ExtrudedTriangle}
    \end{subfigure}
    ~ 
    \begin{subfigure}[t]{0.45\textwidth}
        \centering
        \includegraphics[width=0.485\textwidth]{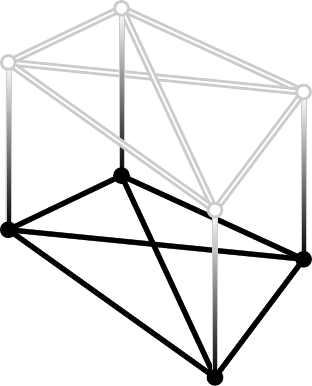}
        \caption{4d8n}
        \label{fig:ExtrudedTetrahedron}
    \end{subfigure}
    \newline 
    \begin{subfigure}[t]{0.45\textwidth}
        \centering
        \includegraphics[width=0.485\textwidth]{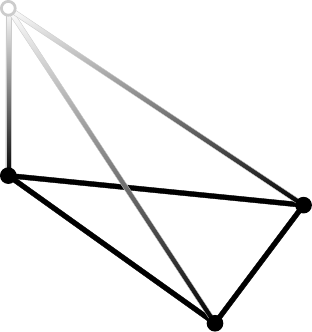}
        \caption{3d4n}
        \label{fig:Tetrahedron}
    \end{subfigure}
    ~ 
    \begin{subfigure}[t]{0.45\textwidth}
        \centering
        \includegraphics[width=0.485\textwidth]{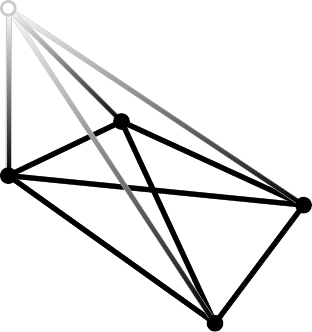}
        \caption{4d5n}
        \label{fig:Pentatope}
    \end{subfigure}
    \caption{Comparison of prism- (top row) and simplex-type (bottom) space-time elements. Black nodes correspond to \(t_n\) and white nodes correspond to \(t_{n+1}\).}
    \label{fig:SimplexElements}
\end{figure}

To start with, we need a spatial mesh in \(n_{sd}\) dimensions, which can be generated by any of the freely or commercially available mesh generators. We confine ourselves to \(n_{sd}\)-simplex-based meshes, such as triangular meshes in 2D and tetrahedral meshes in 3D. In the traditional space-time implementation, the spatial mesh is at first extruded in the time dimension to fill the space-time slab contained between time levels \(t_n\) and \(t_{n+1}\). The extruded mesh is composed of prisms (6-noded space-time elements for 2D problems and 8-noded space-time elements for 3D problems). These elements, referred to as 3d6n and 4d8n in \cite{behr2008simplex} and shown in Figures \ref{fig:ExtrudedTriangle} and \ref{fig:ExtrudedTetrahedron}, respectively, are the basis of the traditional space-time approach and are considered here as a reference.

Our goal is to subdivide these prism-type elements into simplex-type elements, referred to as 3d4n (familiar tetrahedrons) and 4d5n (pentatopes) in \cite{behr2008simplex} and illustrated in \ref{fig:Tetrahedron} and \ref{fig:Pentatope}, respectively. The initial space-time mesh contains only two nodes for each of the nodes in the spatial mesh, one located at the bottom of the slab and the other one located at the top of the slab. The temporal refinement is accomplished by adding, in parts of the domain where temporal accuracy is to be increased, one or more nodes along the line connecting the original nodes in the temporal direction. The space-time faces of the prism-type elements will be later divided into (\(n_{sd}-1\))-node simplices according to the Delaunay criteria, independently for each prism. However, we need to ensure the uniqueness of the Delaunay process and the compatibility of the (\(n_{sd}-1\))-simplices between the neighboring space-time prisms. A solution proposed in \cite{behr2008simplex} for the aforementioned problem is to perturb the time coordinates of some or all nodes randomly. Additionally, \textit{a posteriori} sliver elimination is unavoidable but easily executed by computing 3D or 4D element volumes and by rejecting elements whose volume falls below a specified threshold.
\begin{figure}[!htb]
    \centering
    \includegraphics[width=0.6\textwidth]{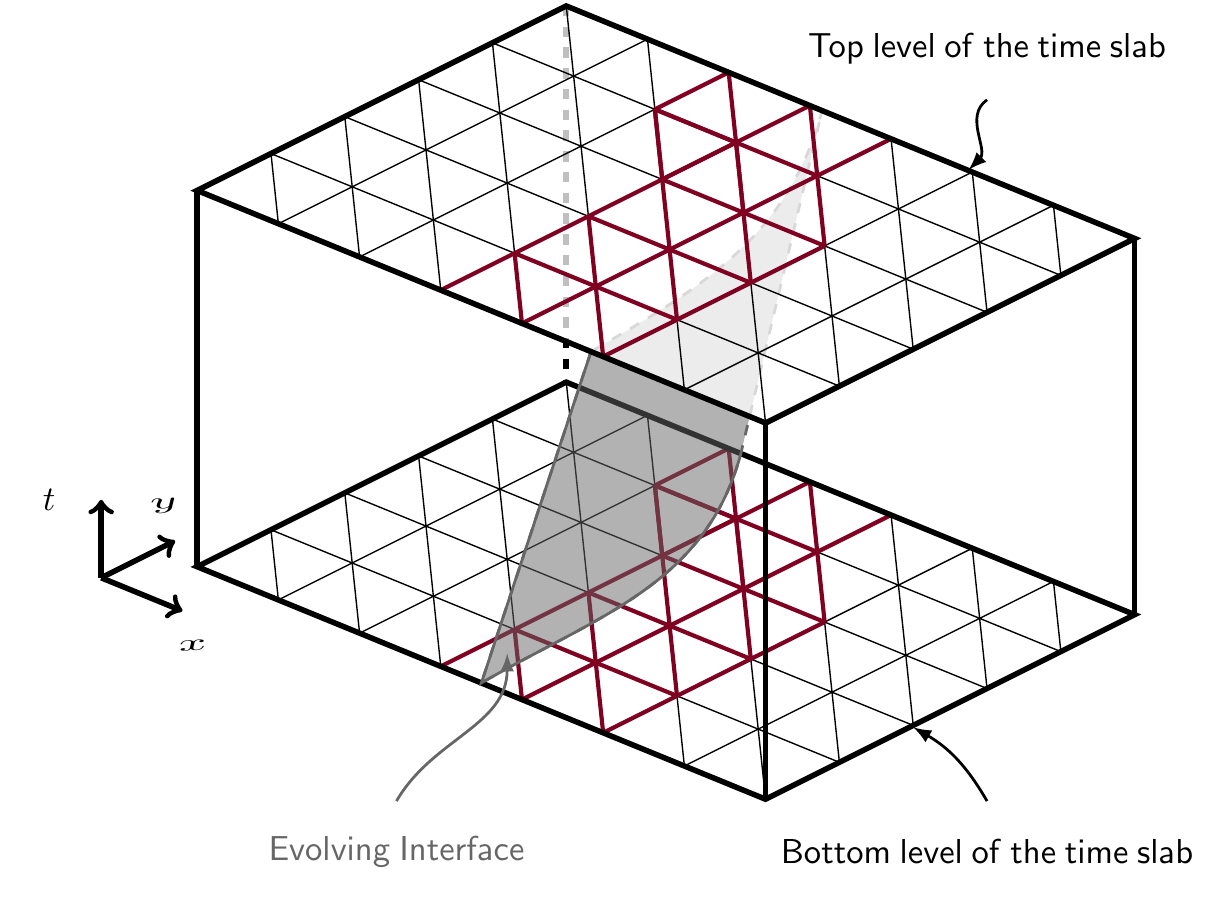}
    \caption{Mesh refinement at a narrow band around the evolving interface. The dark-red-colored elements depict the narrow band.}
    \label{fig:TimeSlab}
\end{figure}

\begin{figure}[!htb]
    \centering
    \includegraphics[width=0.25\textwidth]{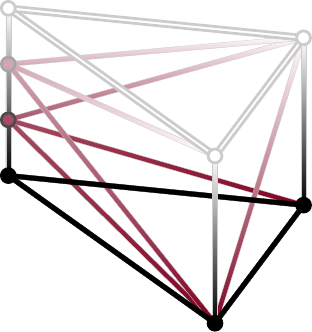}
    \caption{Sample connectivity of a refined element for \(2\)D problems.}
    \label{fig:RefinedElement}
\end{figure}

Considering two-phase flow problems, the region of the domain that demands higher temporal accuracy is the propagating interface between the two fluids. The level-set function gives us the opportunity to refine the mesh at a narrow band around the front, i.e., the zero level-set field, as presented in Figure \ref{fig:TimeSlab}. Inside this narrow band, the temporal refinement can be either uniform or non-uniform. That means that the number of nodes, added along the edges connecting the original nodes in the temporal direction of the extruded elements, can be different among these edges. Furthermore, there are elements outside the narrow band, but still next to it, for which, some of their edges contain more than two nodes in the temporal direction. These elements comprise a transition zone between the refined narrow band and the unrefined region of the domain. The sample connectivity of these elements for \(2\)D problems is illustrated in Figure \ref{fig:RefinedElement}.
\section{Numerical results}
\label{NumericalResults}
\subsection{Rising bubble in 2D}
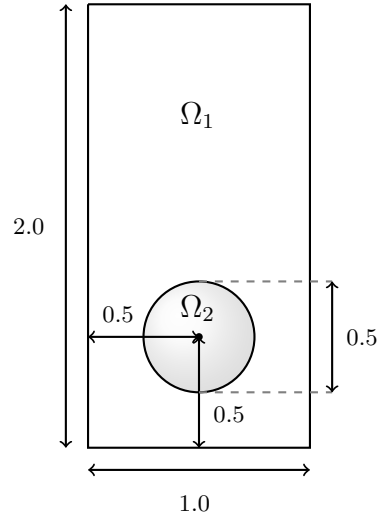
\begin{figure}[!htb]
    \centering
    \begin{tikzpicture}[thick, scale = 2.95,dot/.style={circle,inner sep=1pt,fill}]
    \pgfmathsetmacro{\x}{1}    
    \pgfmathsetmacro{\y}{2}    
    \path (0,\y)  coordinate (A)  (\x,\y)  coordinate (B)
              (\x,0)  coordinate (C)  (0,0)   coordinate (D);
     \draw (A) -- (B) -- (C) -- (D) -- (A);
     \shade[ball color=black!10!white,opacity=0.20] (0.5,0.5) circle (0.25cm);
     \draw (0.5,0.5) circle (0.25cm);
     \node[text width=0.5cm] at (0.5,1.5) {\large \(\Omega_1\)};
    \node[text width=0.5cm] at (0.5,0.63) {\large \(\Omega_2\)};
    \draw[<->] (0,0.5)--(0.5,0.5);
    \node[text width=0.5cm] at (0.15,0.6) {\footnotesize \(0.5\)};
    \draw[<->] (0.5,0)--(0.5,0.5);
    \node[dot] at (0.5,0.5) {};
    \node[text width=0.5cm] at (0.65,0.15) {\footnotesize \(0.5\)};
    \draw[gray,dashed] (0.5,0.75)--(1.1,0.75);
    \draw[gray,dashed] (0.5,0.25)--(1.1,0.25);
    \draw[<->] (1.1,0.25)--(1.1,0.75);
    \node[text width=0.5cm] at (1.25,0.5) {\footnotesize \(0.5\)};
    \draw[<->] (-0.1,0)--(-0.1,2);
    \node[text width=0.5cm] at (-0.25,1.0) {\footnotesize \(2.0\)};
    \draw[<->] (0,-0.1)--(1,-0.1);
    \node[text width=0.5cm] at (0.5,-0.25) {\footnotesize \(1.0\)};
    \end{tikzpicture}
    \captionsetup{justification=centering}
    \caption{Rising bubble in 2D: Computational domain.} \label{fig:RisingBubble2DDomain}
\end{figure}

As a first test case, a two-dimensional bubble, rising in an initially motionless liquid column due to the buoyancy effects, is considered and used for validating the unstructured space-time mesh solver for problems in two-space dimensions. The computational domain is rectangular and has the size of \(\SI{1.0}{m} \times \SI{2.0}{m}\). An initially circular bubble with diameter \(d = \SI{0.5}{m}\) is placed inside the domain, as illustrated in Figure \ref{fig:RisingBubble2DDomain}.

The fluids have the following properties: \(\rho_1 = \SI{1000}{kg/m^3}\), \(\rho_2 = \SI{100}{kg/m^3}\), \(\mu_1 = \SI{10}{kg/m/s}\) and \(\mu_2 = \SI{1}{kg/m/s}\). The applied gravitational acceleration is \(f_y  = -g = \SI{-0.98}{m/s^2}\) and the surface tension coefficient \(\gamma\) is equal to \(\SI{24.5}{kg/s^2}\). The characteristic dimensionless Reynolds and E\"otv\"os numbers are the following:
\begin{equation} \label{Re_RisingBubble}
Re = \frac{\rho_2\sqrt{gd}d}{\mu_2} = 35,
\end{equation}
\begin{equation} \label{Eo_RisingBubble}
Eo = \frac{g\rho_2d^2}{\gamma} = 10.
\end{equation}
\begin{figure}[!htb]
    \centering
    \includegraphics[width=1.0\textwidth]{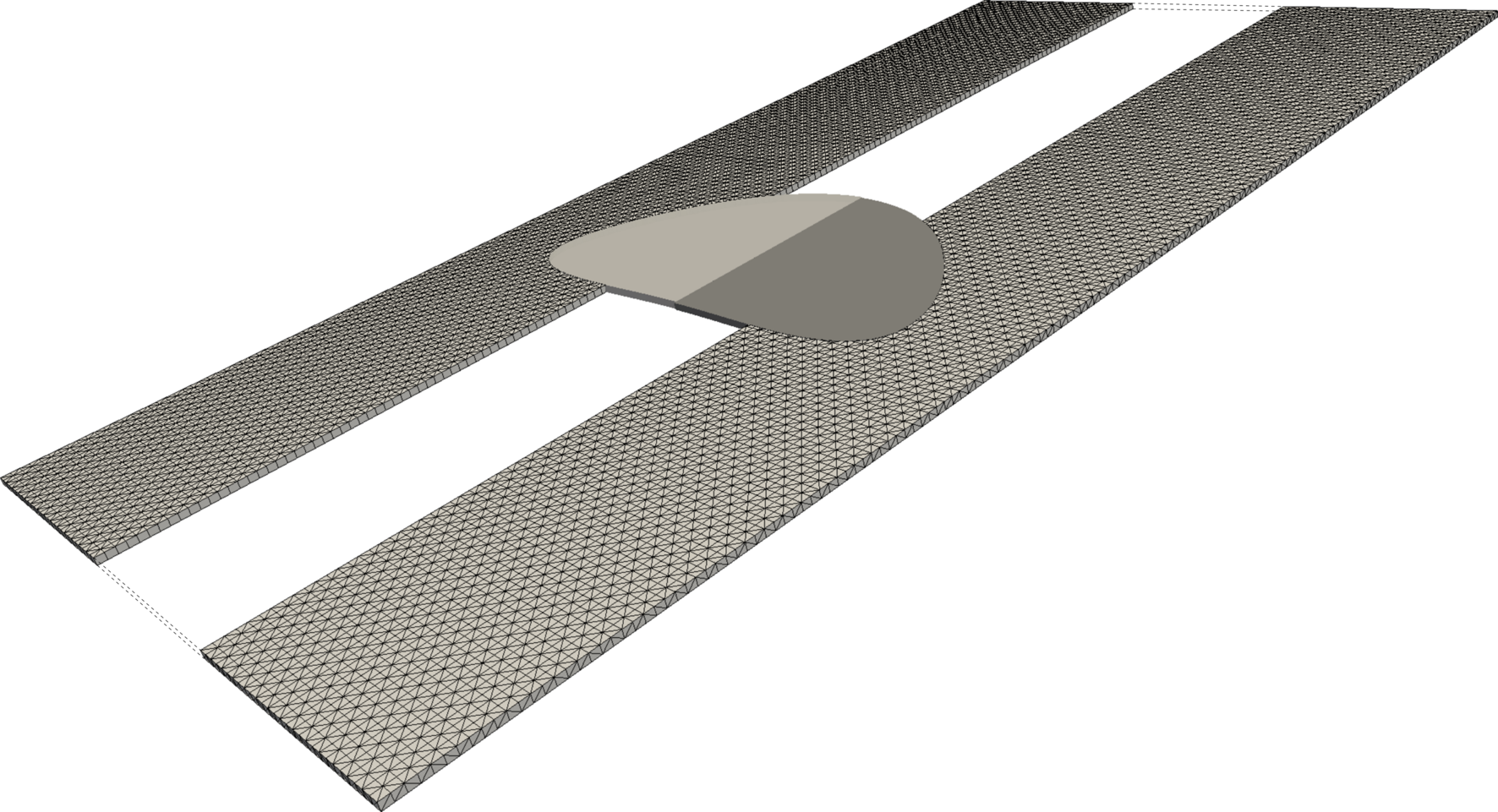}
    \caption{Space-time discretization corresponding to one of the \(\SI{300}{}\) (in total) time slabs of the simulation. Light grey color corresponds to the prismatic space-time discretization (back half of the bubble), and dark grey color corresponds to simplex-based space-time discretization (front half of the bubble).}\label{fig:RisingBubble2DTimeSlab}
\end{figure}
The spatial resolution consists of triangular elements. The aforementioned triangulation is based on an equidistant grid of \(\SI{80}{} \times \SI{160}{}\) quadrilateral elements. The time-slab size is \(\Delta t = \SI{0.01}{s}\). No-slip boundary condition is imposed on the top and bottom of the computational domain and slip boundary condition on the vertical walls. Zero pressure is defined at the upper wall, and the initial velocity field is set to zero. With the given configuration, the bubble shape should become ellipsoidal, as stated in \cite{hysing2009quantitative}. That means, the surface tension effects are dominant enough to hold the bubble together and no break up should be expected.

\begin{figure}[!htb]
    \centering
    \begin{subfigure}[t]{0.3\textwidth}
        \includegraphics[width=\textwidth]{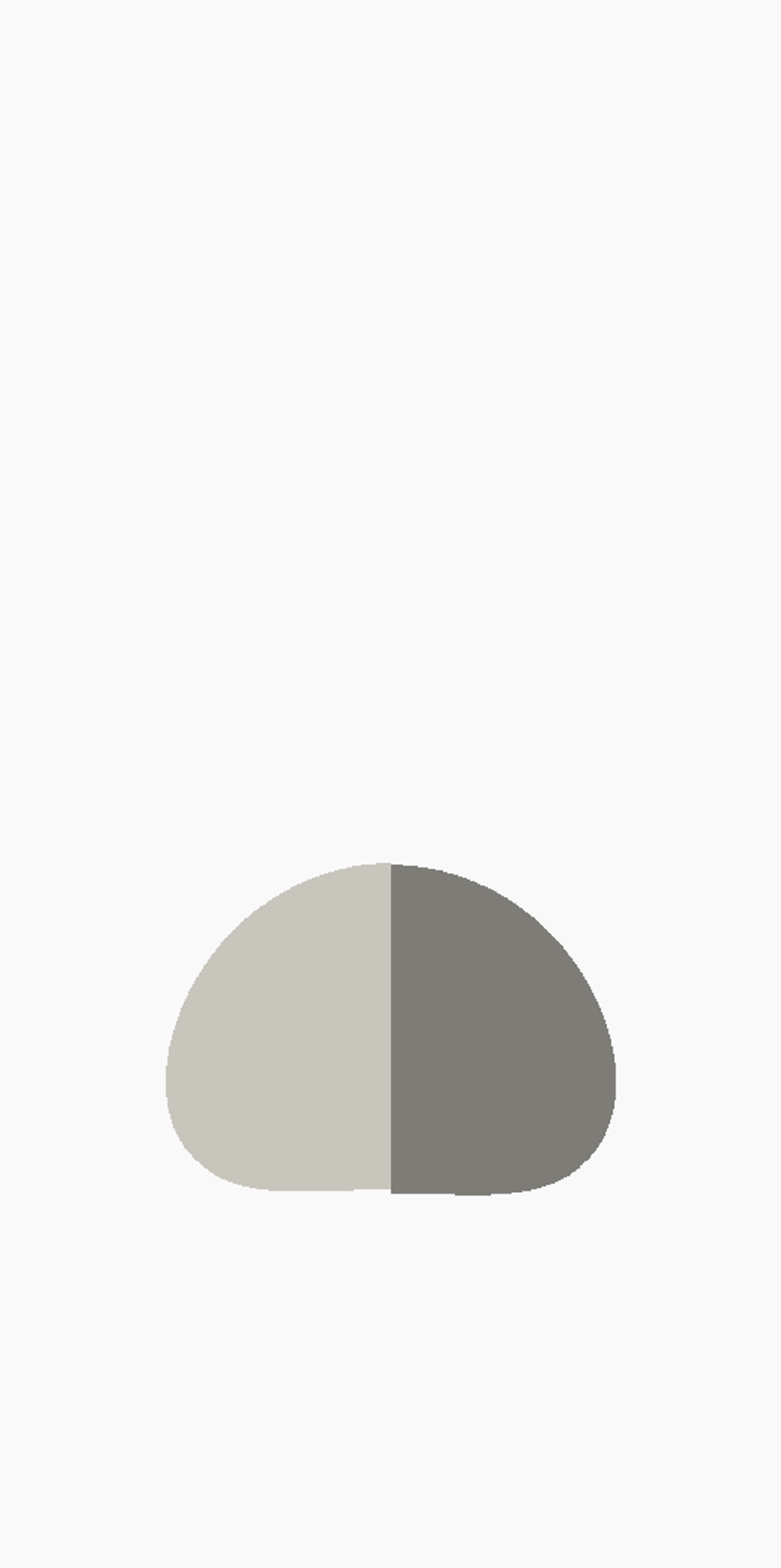}
        \caption{\(t= \SI{1.0}{s}\)}
        \label{fig:RisingBubble2D1s}
    \end{subfigure}
    ~ 
    \begin{subfigure}[t]{0.3\textwidth}
        \includegraphics[width=\textwidth]{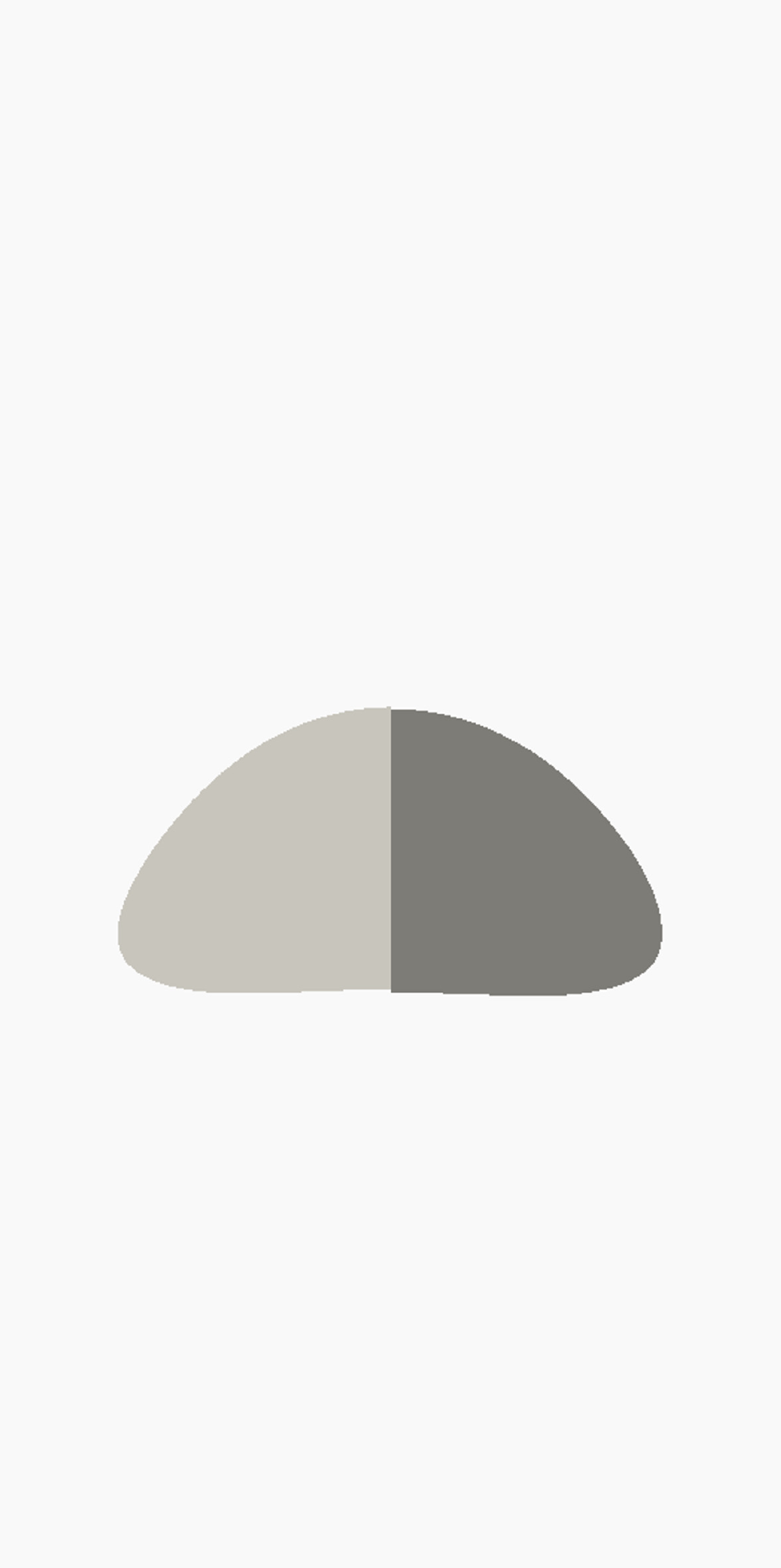}
        \caption{\(t= \SI{2.0}{s}\)}
        \label{fig:RisingBubble2D2s}
    \end{subfigure}
    ~ 
    \begin{subfigure}[t]{0.3\textwidth}
        \includegraphics[width=\textwidth]{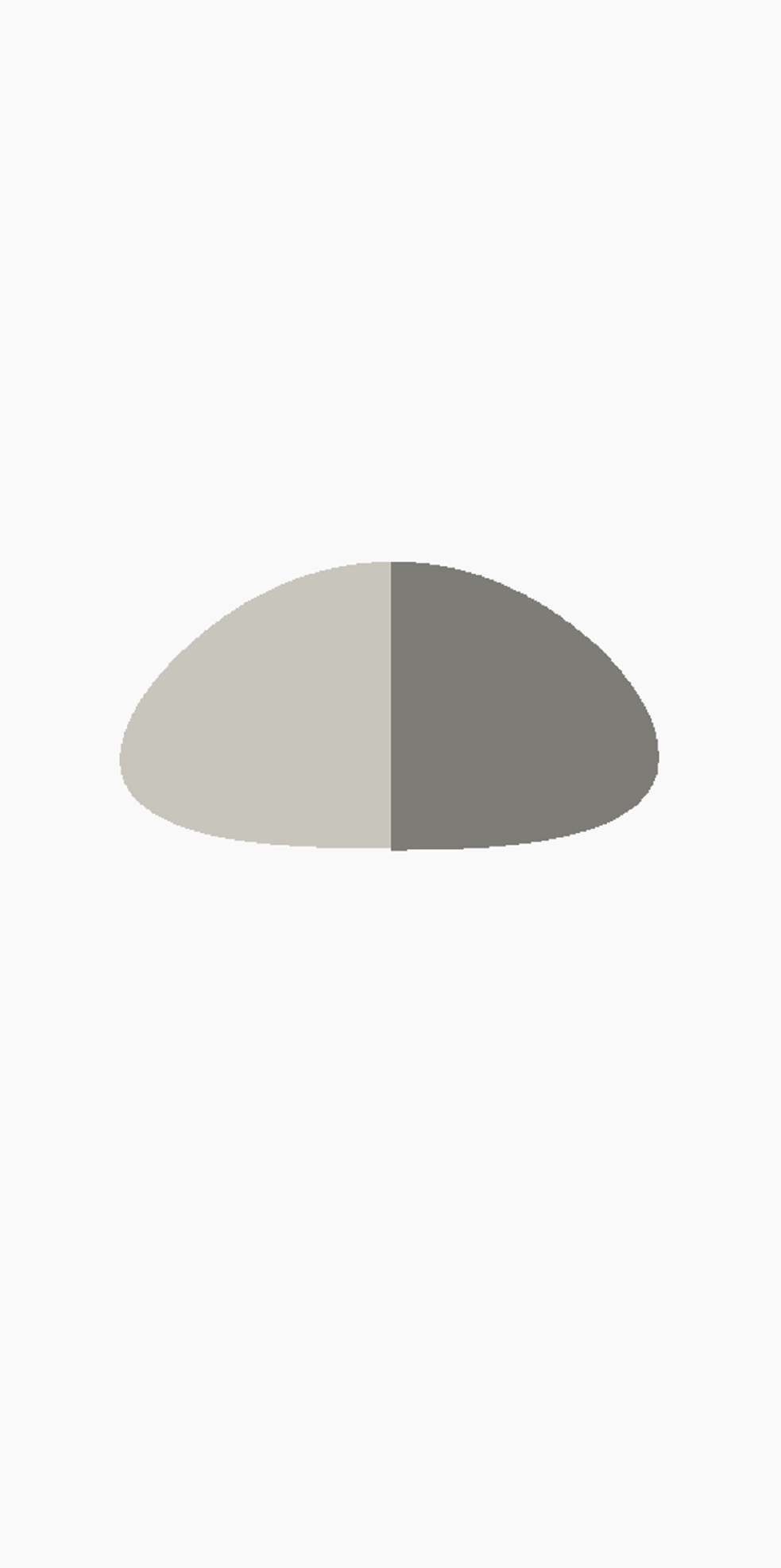}
        \caption{\(t= \SI{3.0}{s}\)}
        \label{fig:RisingBubble2D3s}
    \end{subfigure}
    \caption{Bubble position at various time instances, obtained using a prism-type space-time discretization and a simplex-type space-time discretization. Light grey color corresponds to the prismatic space-time discretization (left half of the bubble) and dark grey color corresponds to simplex-based space-time discretization (right half of the bubble).}\label{fig:RisingBubble2DShape}
\end{figure}

The bubble rose first for \(t = \SI{3.0}{s}\) using the usual discontinuous-in-time Galerkin time stepping (prismatic space-time elements, as shown in Figure \ref{fig:RisingBubble2DTimeSlab}).  These standard results were then compared with the results obtained with the tetrahedral-based space-time mesh discretization of a slab of the same thickness without applying any temporal refinement (cf. Figure \ref{fig:RisingBubble2DTimeSlab}). Figure \ref{fig:RisingBubble2DShape} shows the rising bubble over time and compares the bubble position obtained using the aforementioned types of discretization.
\begin{figure}[!htb]
    \includegraphics[width=\textwidth]{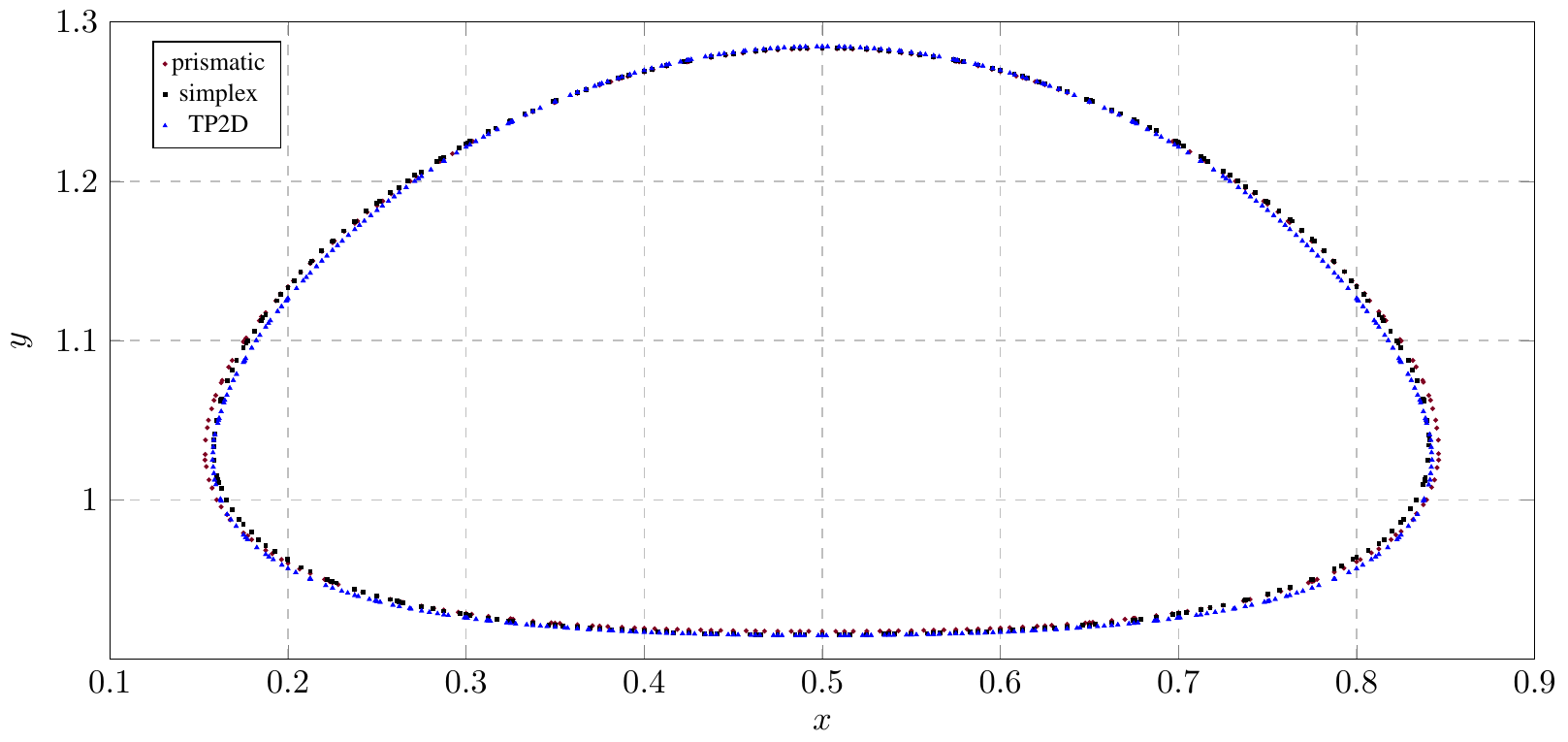}
    \caption{Comparison of the bubble at \(t= \SI{3.0}{s}\) with reference data published by \citet{hysing2009quantitative}.}
    \label{fig:RisingBubble2DShapeComparison}
\end{figure}
\begin{figure}[!htb]
    \centering
    \begin{subfigure}[t]{0.475\textwidth}
    	\centering
        \includegraphics[width=0.83\textwidth]{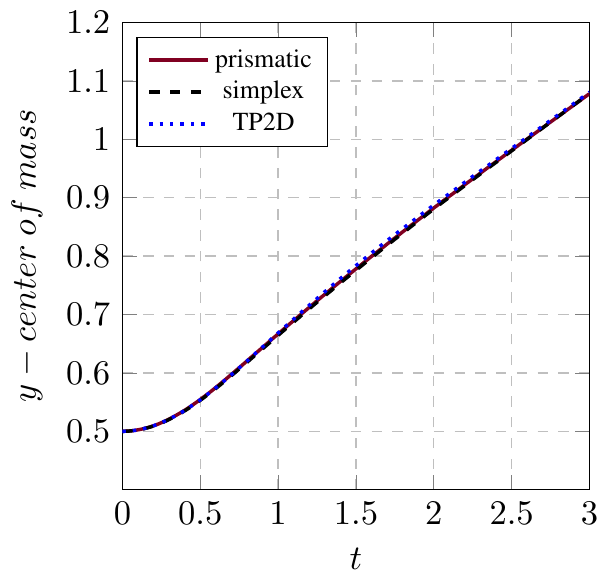}
        \caption{Over the whole simulation time.}
        \label{fig:RisingBubble2DCenterOfMassfull}
    \end{subfigure}
    ~ 
    \begin{subfigure}[t]{0.475\textwidth}
    	\centering
        \includegraphics[width=0.85\textwidth]{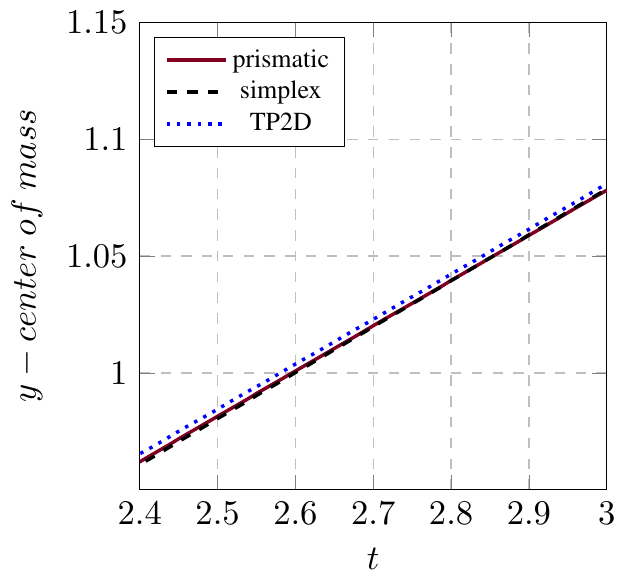}
        \caption{Between the time instances \(t_1= \SI{2.4}{s}\) and \(t_2= \SI{3.0}{s}\).}
        \label{fig:RisingBubble2DCenterOfMasszoom}
    \end{subfigure}
    \caption{The position of the center of mass \(\textbf{X}_c\) in \(y\)-direction of the rising bubble in 2D.}
    \label{fig:RisingBubble2DCenterOfMass}
\end{figure}

In Figure \ref{fig:RisingBubble2DShapeComparison} the bubble shape at \(t = \SI{3.0}{s}\) is compared with results published in \cite{hysing2009quantitative}. As we can see, the results obtained with both types of the different space-time elements show a very good agreement with the reference data. The centroid of the two-dimensional bubble, which is defined by:
\begin{equation} \label{Centroid_RisingBubble}
\bold{X}_c =(x_c, y_c) = \frac{\int_{\Omega_2}\bold{x} \ dx}{\int_{\Omega_2}1 \ dx},
\end{equation}
is also tracked.

Figure \ref{fig:RisingBubble2DCenterOfMass} depicts the position of the center of mass of the bubble. This quantity is also in good agreement with the TP2D simulation results reported in \cite{hysing2009quantitative}. We have compared our results only with TP2D simulation results, published in \cite{hysing2009quantitative}, in order that the visibility of the Figures \ref{fig:RisingBubble2DShapeComparison} and \ref{fig:RisingBubble2DCenterOfMass} could be maintained. All the different approaches in \cite{hysing2009quantitative} lead to much the same results for this test case as well.
\subsection{Rising droplet in 3D}
The benchmark case of the rising bubble in 2D is now repeated in 3D. This example serves as the initial validation of the unstructured space-time mesh solver for problems in three-space dimensions. The bubble rose first for \(t = \SI{3.0}{s}\) using the usual discontinuous-in-time Galerkin time stepping (prismatic space-time elements, linear-in-time interpolation). The time step size was at first chosen as \(\Delta t = \SI{0.01}{s}\). These standard results were then compared with the rising results obtained with the pentatope-based space-time mesh discretization of a slab \(\Delta t = \SI{0.01}{s}\) thick without any temporal refinement.
\begin{figure}[!htb]
    \centering
    \begin{tikzpicture}[thick, scale = 2.95,dot/.style={circle,inner sep=1pt,fill}]
        \pgfmathsetmacro{\x}{1}    
    \pgfmathsetmacro{\y}{1}    
    \pgfmathsetmacro{\z}{2}    
    \path (0,0,\y)  coordinate (A)  (\x,0,\y)  coordinate (B)
         (\x,0,0)  coordinate (C)  (0,0,0)   coordinate (D)
         (0,\z,\y) coordinate (E)  (\x,\z,\y) coordinate (F)
         (\x,\z,0) coordinate (G)  (0,\z,0)  coordinate (H);
    \draw (A) -- (B) -- (C) -- (G) -- (F) -- (B) (A)--(E)--(F)--(G)--(H) --(E);
    \draw[gray,dashed]  (A) -- (D) -- (C)   (D) -- (H);
    \draw (0.25,0.5,0.5) arc (180:360:0.25cm and 0.125cm);
    \shade[ball color=black!10!white,opacity=0.20] (0.5,0.5,0.5) circle (0.25cm);
    \draw (0.5,0.5,0.5) circle (0.25cm);
    \node[text width=0.5cm] at (0.5,1.5,0.5) {\large \(\Omega_1\)};
    \node[text width=0.5cm] at (0.5,0.63,0.5) {\large \(\Omega_2\)};
    \draw[<->] (0,0,0.5)--(0.5,0,0.5);
    \node[text width=0.5cm] at (0.25,-0.05,0.6) {\footnotesize \(0.5\)};
    \draw[<->] (0.5,0,1.0)--(0.5,0,0.5);
    \node[text width=0.5cm] at (0.625,-0.05,0.7) {\footnotesize \(0.5\)};
    \draw[<->] (0.5,0,0.5)--(0.5,0.5,0.5);
    \node[dot] at (0.5,0.5,0.5) {};
    \node[text width=0.5cm] at (0.65,0.1,0.5) {\footnotesize \(0.5\)};
    \draw[gray,dashed] (0.5,0.75,0.5)--(1.35,0.75,0.5);
    \draw[gray,dashed] (0.5,0.25,0.5)--(1.35,0.25,0.5);
    \draw[<->] (1.35,0.25,0.5)--(1.35,0.75,0.5);
    \node[text width=0.5cm] at (1.50,0.5,0.5) {\footnotesize \(0.5\)};
    \draw[<->] (-0.1,0,1.0)--(-0.1,2,1.0);
    \node[text width=0.5cm] at (-0.25,1.0,1.0) {\footnotesize \(2.0\)};
    \draw[<->] (0,-0.1,1.0)--(1,-0.1,1.0);
    \node[text width=0.5cm] at (0.5,-0.25,1.0) {\footnotesize \(1.0\)};
    \end{tikzpicture}
    \captionsetup{justification=centering}
    \caption{Rising droplet in 3D: Computational domain.} \label{fig:RisingBubble3DDomain}
\end{figure}
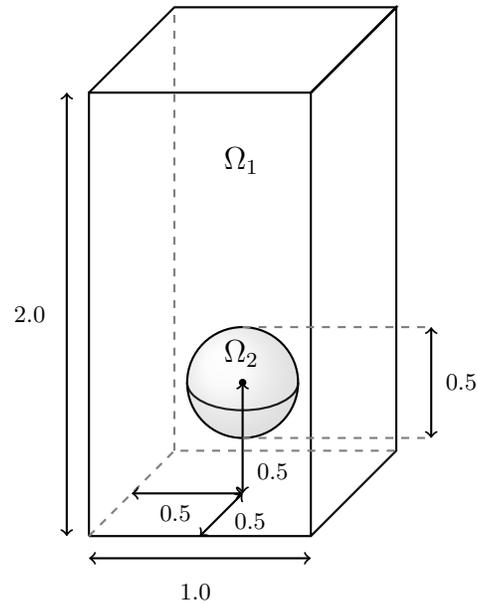

The computational domain occupies a cuboid tank with the dimensions \(\SI{1.0}{m} \times \SI{2.0}{m} \times \SI{1.0}{m}\). The initial position of the droplet is at the center of the tank at a distance of \(\SI{0.5}{m}\) from the bottom wall and its initial shape is assumed to be spherical (cf. Figure \ref{fig:RisingBubble3DDomain}). The properties of the fluids are given by: \(\rho_1 = \SI{1000}{kg/m^3}\), \(\rho_2 = \SI{100}{kg/m^3}\), \(\mu_1 = \SI{10}{kg/m/s}\), \(\mu_2 = \SI{1}{kg/m/s}\) and \(f_y  = -g = \SI{-0.98}{m/s^2}\). The surface tension coefficient is \(\gamma = \SI{24.5}{kg/s^2}\). No-slip conditions are applied on all outer boundaries. Zero pressure is applied at the top wall and the initial velocity field is set to zero. Due to buoyancy effects, the droplet will start rising, change its shape and become ellipsoidal-shaped.
\begin{figure}[!htb]
    \centering
    \begin{subfigure}[t]{0.3\textwidth}
        \includegraphics[width=\textwidth]{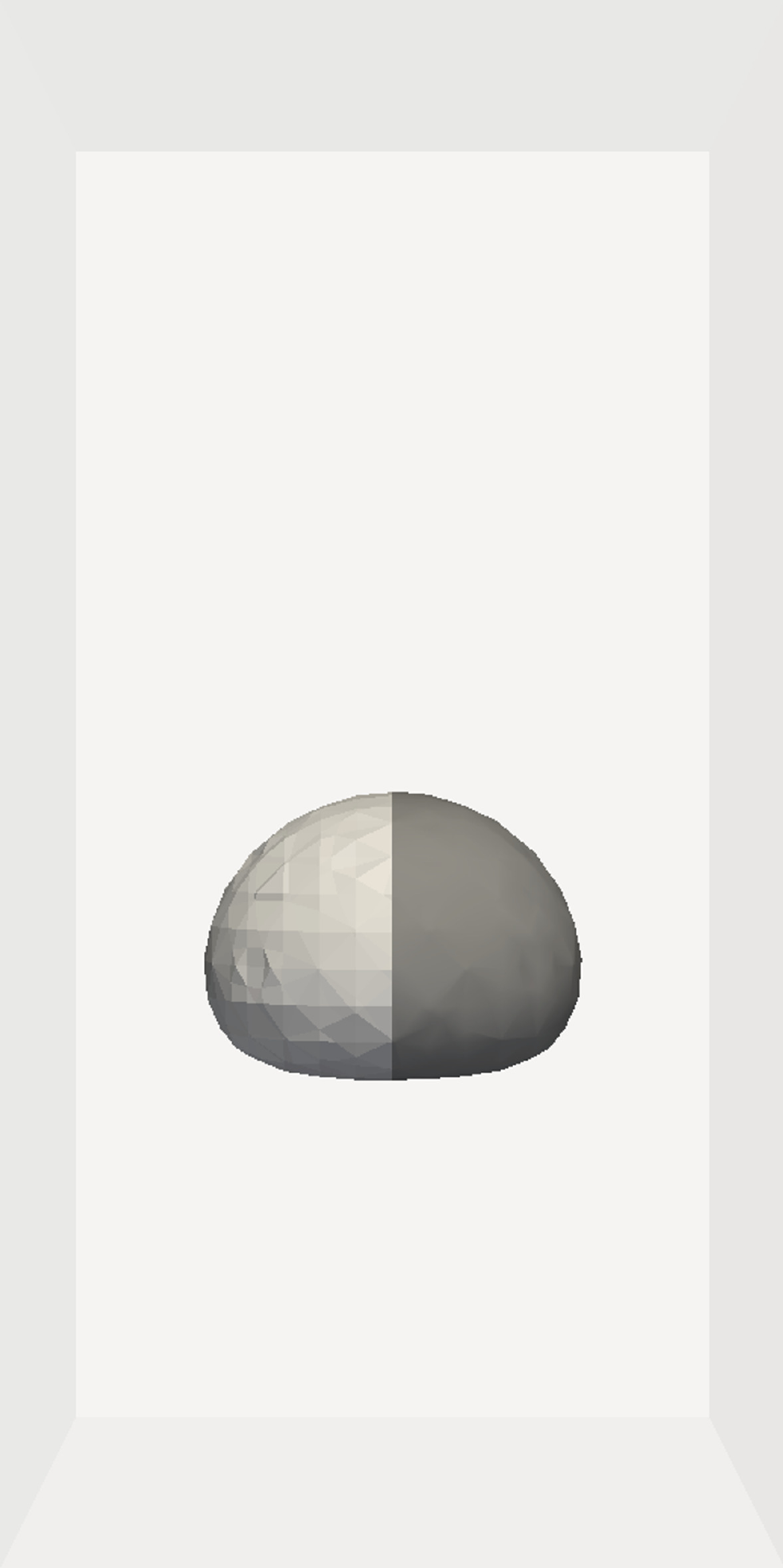}
        \caption{\(t= \SI{1.0}{s}\)}
        \label{fig:RisingBubble3D1s}
    \end{subfigure}
    ~ 
    \begin{subfigure}[t]{0.3\textwidth}
        \includegraphics[width=\textwidth]{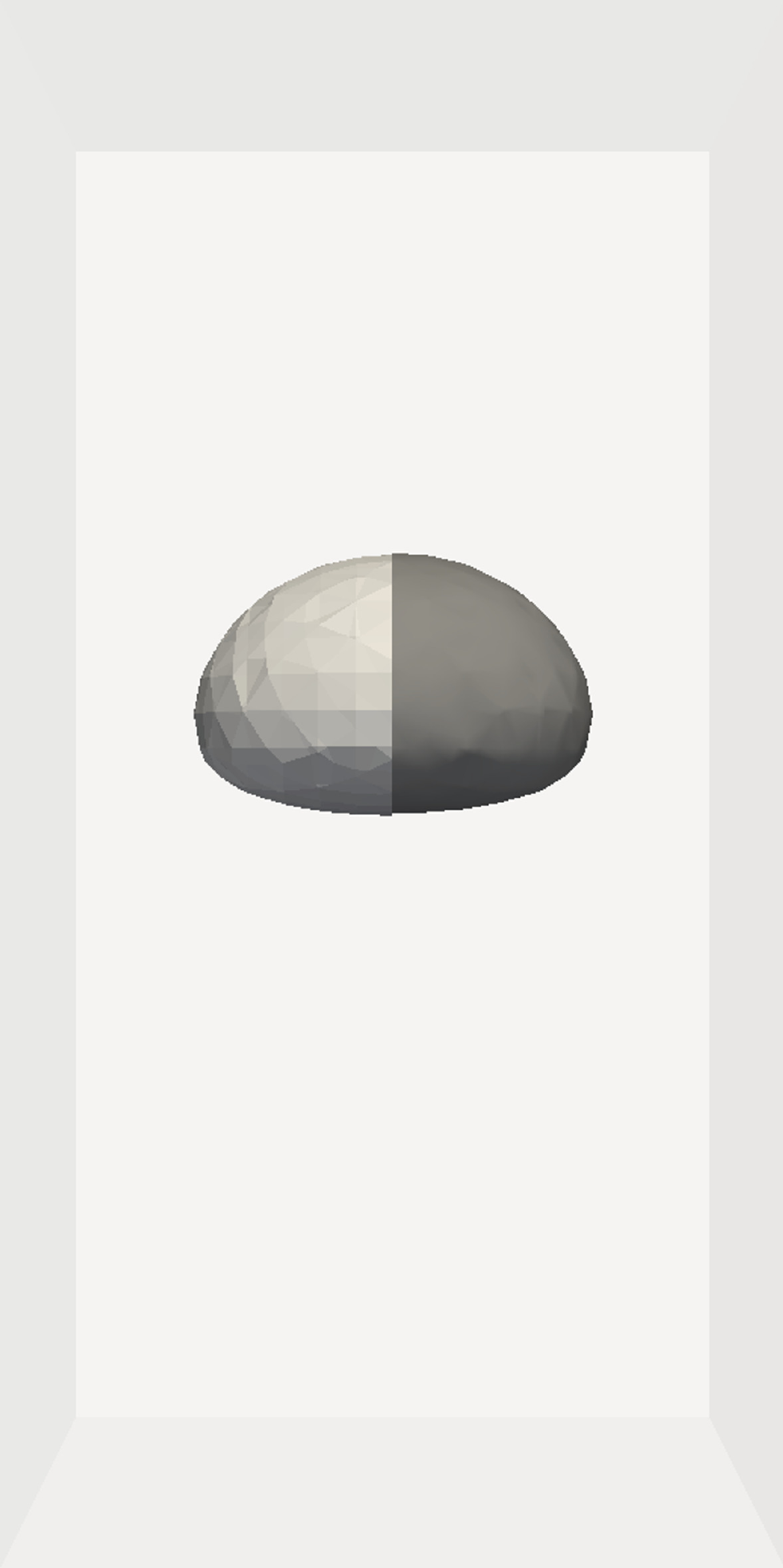}
        \caption{\(t= \SI{2.0}{s}\)}
        \label{fig:RisingBubble3D2s}
    \end{subfigure}
    ~ 
    \begin{subfigure}[t]{0.3\textwidth}
        \includegraphics[width=\textwidth]{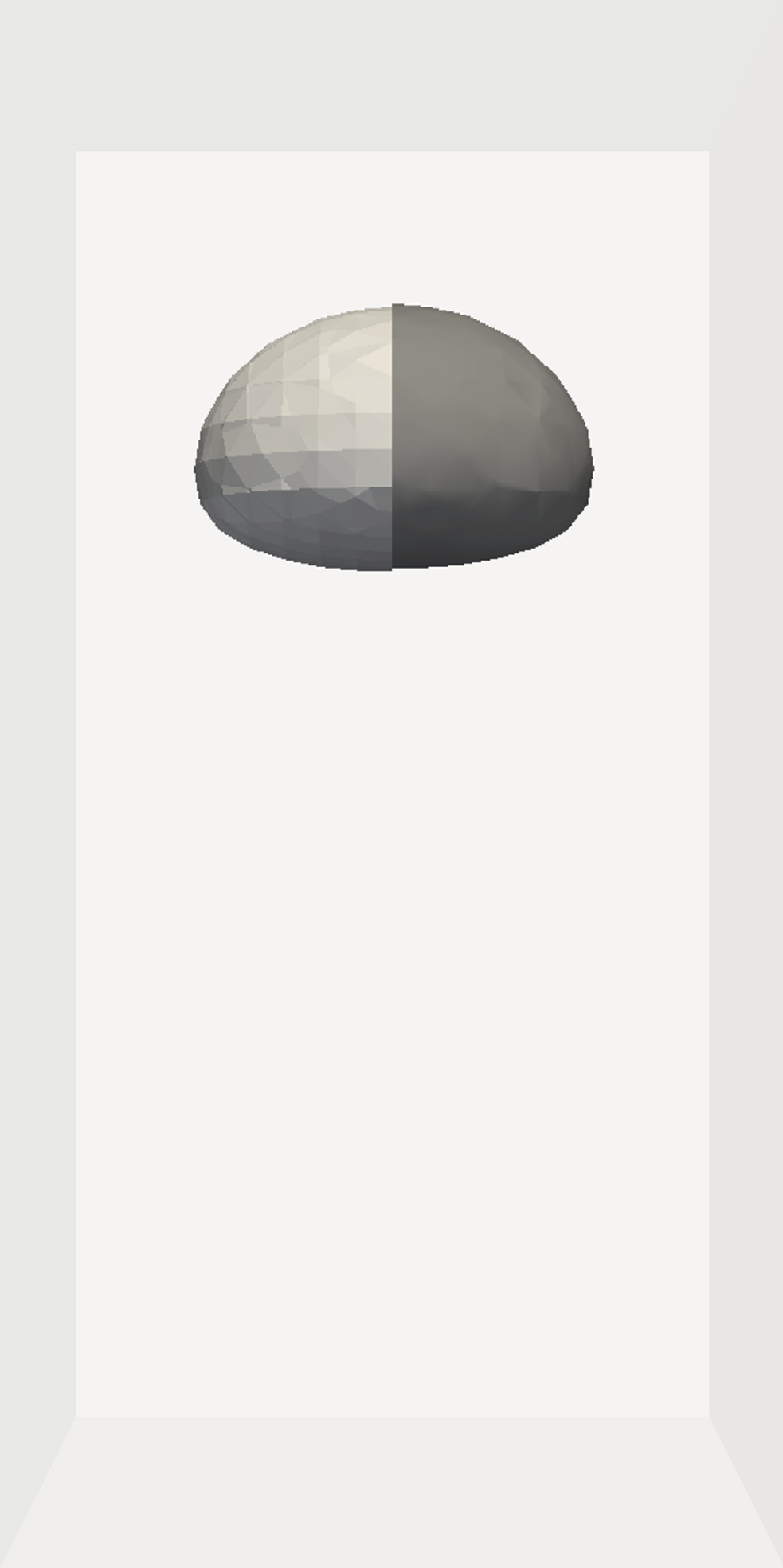}
        \caption{\(t= \SI{3.0}{s}\)}
        \label{fig:RisingBubble3D3s}
    \end{subfigure}
    \caption{Droplet position at various time instances, obtained using a prism-type space-time discretization and a simplex-type space-time discretization. Light grey color corresponds to the prismatic space-time discretization (left half of the droplet) and dark grey color corresponds to simplex-based space-time discretization (right half of the droplet).}\label{fig:RisingBubble3DShape}
\end{figure}

\begin{figure}[!htb]
    \centering
    \begin{subfigure}[t]{0.475\textwidth}
        \includegraphics[width=0.83\textwidth]{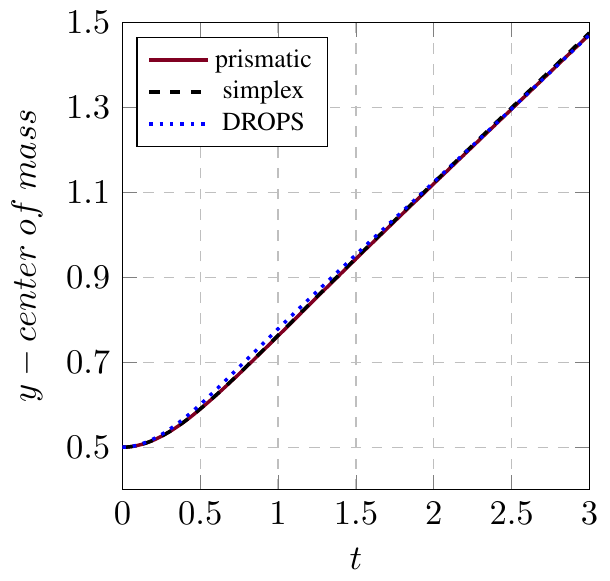}
        \caption{Over the whole simulation time.}
        \label{fig:RisingDroplet3DCenterOfMassfull}
    \end{subfigure}
    ~ 
    \begin{subfigure}[t]{0.475\textwidth}
        \includegraphics[width=0.85\textwidth]{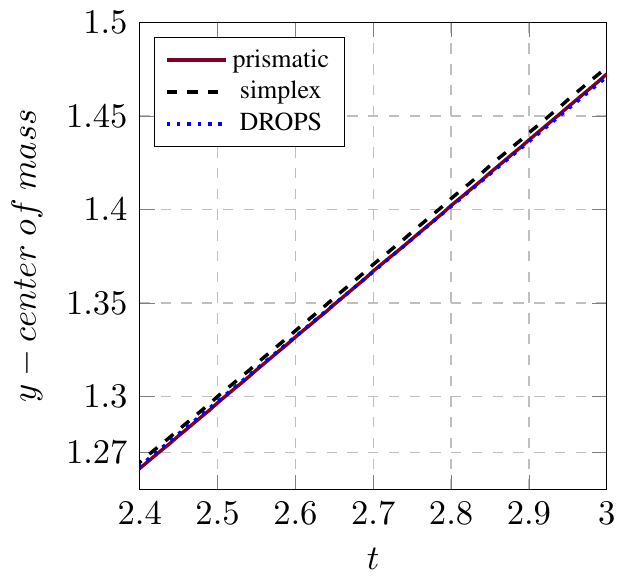}
        \caption{Between the time instances \(t_1= \SI{2.4}{s}\) and \(t_2= \SI{3.0}{s}\).}
        \label{fig:RisingDroplet3DCenterOfMasszoom}
    \end{subfigure}
    \caption{The position of the center of mass \(\textbf{X}_c\) in \(y\)-direction of the rising droplet in 3D.}
    \label{fig:RisingDroplet3DCenterOfMass}
\end{figure}

Figure \ref{fig:RisingBubble3DShape} shows the rising droplet over time and compares the simulation results of the droplet position obtained with the prismatic and the simplex space-time mesh discretization. We also track the position of the centroid of the three-dimensional droplet, in order to acquire a more accurate comparison. The center of the mass is defined by: 
\begin{equation} \label{Centroid_RisingDroplet}
\bold{X}_c =(x_c, y_c, z_c) = \frac{\int_{\Omega_2}\bold{x} \ dx}{\int_{\Omega_2}1 \ dx}.
\end{equation}
The position of the center of mass of the droplet is presented in Figure \ref{fig:RisingDroplet3DCenterOfMass}. This quantity is also in good agreement with the DROPS simulation results reported in \cite{adelsberger20143d}. We have compared our results only with DROPS simulation results, published in \cite{adelsberger20143d}, again for maintaining the visibility of the Figure \ref{fig:RisingDroplet3DCenterOfMass}. NaSt3DGPF, reported in \cite{adelsberger20143d}, leads to very similar results for this test case as well. However, the position of the droplet obtained with OpenFOAM is shifted in the vertical direction. One explanation for this difference might be that OpenFOAM employs the VOF method as an interface capturing method, as stated by \citet{adelsberger20143d}.
\subsection{Step cavity  in 2D with time refinement}
\label{StepCavity}
As the last example for verifying our numerical approach, we simulate the filling stage of a step cavity, which has a lower and an upper step, considered by \citet{dhatt1990finite} and \citet{Tezduyar2002}. The molten material enters the mold with uniform velocity and displaces the air, which is initially quiescent. This inflow velocity is deliberately chosen to be low. As a consequence, the spreading impact of gravity on the evolving interface is apparent.
\begin{figure}[!htb]
    \centering
    \begin{tikzpicture}[thick, scale = 2.95,dot/.style={circle,inner sep=1pt,fill}]
        \pgfmathsetmacro{\x}{0.8}    
    \pgfmathsetmacro{\y}{0.3}    
    \path (0,0)                       coordinate (A)  (\x+0.6,0)          coordinate (B)
         (\x+0.6,\y)              coordinate (C)  (\x+0.6+0.6,\y)  coordinate (D)
         (\x+0.6+0.6,\y+\y)  coordinate (E)  (\x,\y+\y)           coordinate (F)
        (\x,\y)                     coordinate (G)  (0,\y)               coordinate (H);
    \draw (A) -- (B) -- (C) -- (D) -- (E) -- (F) -- (G) --  (H) -- (A);
    \filldraw[fill=black!20!white,opacity=0.20, draw=black] (0,0) rectangle (0.2,0.3);
    \node[text width=0.5cm] at (1.1,0.3) {\large \(\Omega_1\)};
    \node[text width=0.5cm] at (0.1,0.15) {\large \(\Omega_2\)};
    \draw[gray,dashed] (0.8,0.3)--(0.8,-0.1);
    \draw[gray,dashed] (2.0,0.3)--(2.0,-0.1);
    \draw[gray,dashed] (1.4,0.0)--(2.1,0.0);
    \draw[-] (-0.1,0.0)--(-0.1,0.3);
    \draw[->] (-0.1,0.0)--(0.0,0.0);
    \draw[->] (-0.1,0.1)--(0.0,0.1);
    \draw[->] (-0.1,0.2)--(0.0,0.2);
    \draw[->] (-0.1,0.3)--(0.0,0.3);
    \draw[-] (1.9,0.3)--(1.9,0.6);
    \draw[->] (1.9,0.3)--(2.0,0.3);
    \draw[->] (1.9,0.4)--(2.0,0.4);
    \draw[->] (1.9,0.5)--(2.0,0.5);
    \draw[->] (1.9,0.6)--(2.0,0.6);
    \draw[<->] (2.1,0.3)--(2.1,0.6);
    \node[text width=0.5cm] at (2.20,0.45) {\footnotesize \(0.3\)};
    \draw[<->] (2.1,0)--(2.1,0.3);
    \node[text width=0.5cm] at (2.20,0.15) {\footnotesize \(0.3\)};
    \draw[<->] (0,-0.1)--(0.8,-0.1);
    \node[text width=0.5cm] at (0.4,-0.20) {\footnotesize \(0.8\)};
    \draw[<->] (0.8,-0.1)--(1.4,-0.1);
    \node[text width=0.5cm] at (1.1,-0.20) {\footnotesize \(0.6\)};
    \draw[<->] (1.4,-0.1)--(2.0,-0.1);
    \node[text width=0.5cm] at (1.7,-0.20) {\footnotesize \(0.6\)};
    \end{tikzpicture}
    \captionsetup{justification=centering}
    \caption{Step Cavity in 2D: Computational domain.} \label{fig:StepCavity2D}
\end{figure}
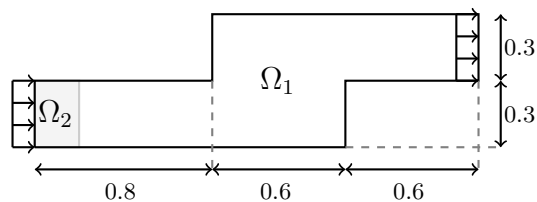

The computational domain is illustrated in Figure \ref{fig:StepCavity2D}. The spatial discretization of the domain consists of \(\SI{365}{}\) triangular elements. The time-slab size is \(\Delta t = \SI{0.005}{s}\). Slip boundary conditions are assumed on the horizontal and vertical walls, except for those of the inflow (leftmost vertical boundary) and outflow (rightmost vertical boundary).  A uniform velocity is imposed at the inflow boundary, whereas traction-free boundary conditions are used at the outflow boundary. We consider isothermal condition, so natural convection and phase-change effects are disregarded.
\begin{figure}[!htb]
    \includegraphics[width=\textwidth]{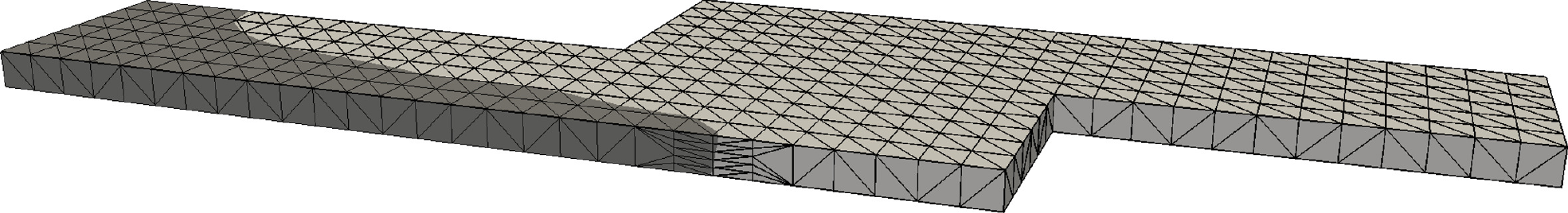}
    \caption{Hybrid space-time mesh corresponding to one of the \(\SI{320}{}\) (in total) time slabs of the simulation.}
    \label{fig:SpaceTimeSlabStepCavity}
\end{figure}

The material properties correspond to those of \cite{dhatt1990finite} and are as follows: \(\rho_1 = \SI{0.1}{kg/m^3}\), \(\rho_2 = \SI{100}{kg/m^3}\), \(\mu_1 = \SI{0.02}{kg/m/s}\), \(\mu_2 = \SI{0.2}{kg/m/s}\). The gravitational acceleration is equal to \(f_y  = -g = \SI{-9.80}{m/s^2}\). The surface tension effects and the wall friction are neglected.
\begin{figure}[!htb]
    \centering
    \begin{subfigure}[t]{0.3\textwidth}
        \includegraphics[width=\textwidth]{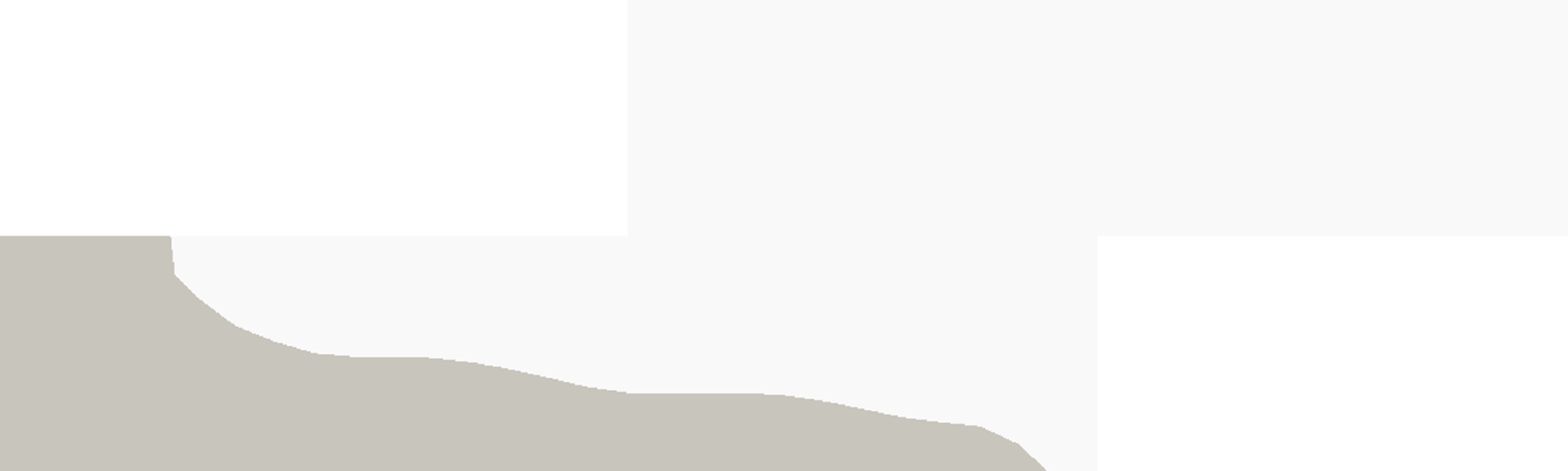}
        \label{fig:StepCavityPrism2D04s}
    \end{subfigure}
    ~ 
    \begin{subfigure}[t]{0.3\textwidth}
        \includegraphics[width=\textwidth]{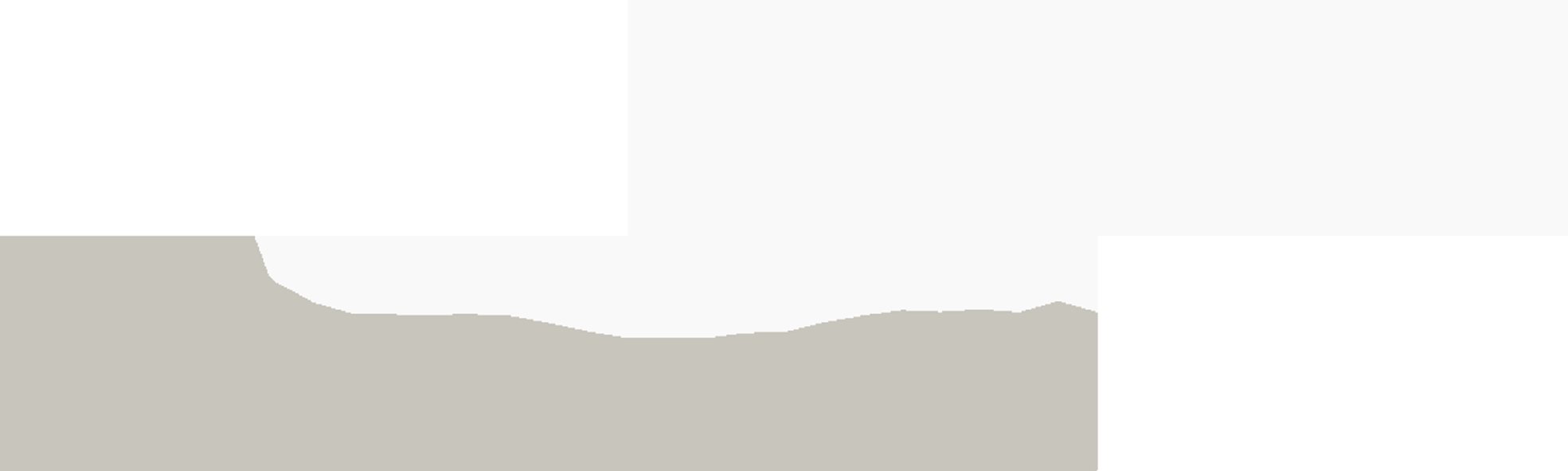}
        \label{fig:StepCavityPrism2D08s}
    \end{subfigure}
    ~ 
    \begin{subfigure}[t]{0.3\textwidth}
        \includegraphics[width=\textwidth]{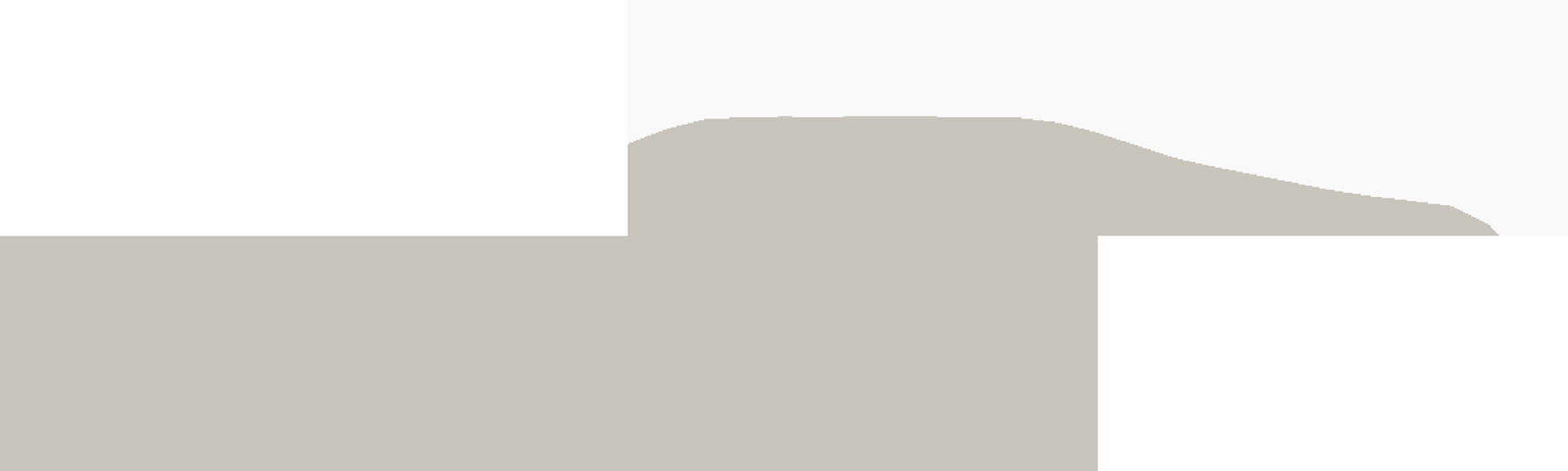}
        \label{fig:StepCavityPrism2D16s}
    \end{subfigure}\\
    \begin{subfigure}[t]{0.3\textwidth}
        \includegraphics[width=\textwidth]{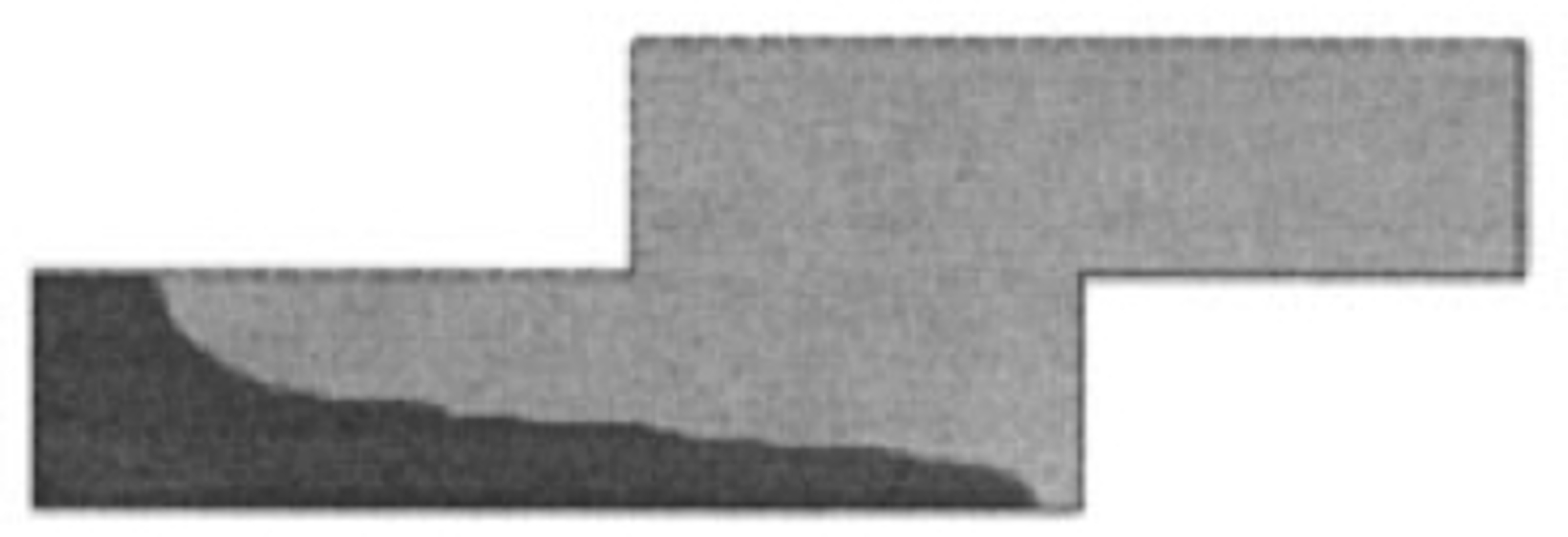}
        \label{fig:StepCavityRef2D04s}
    \end{subfigure}
    ~ 
    \begin{subfigure}[t]{0.3\textwidth}
        \includegraphics[width=\textwidth]{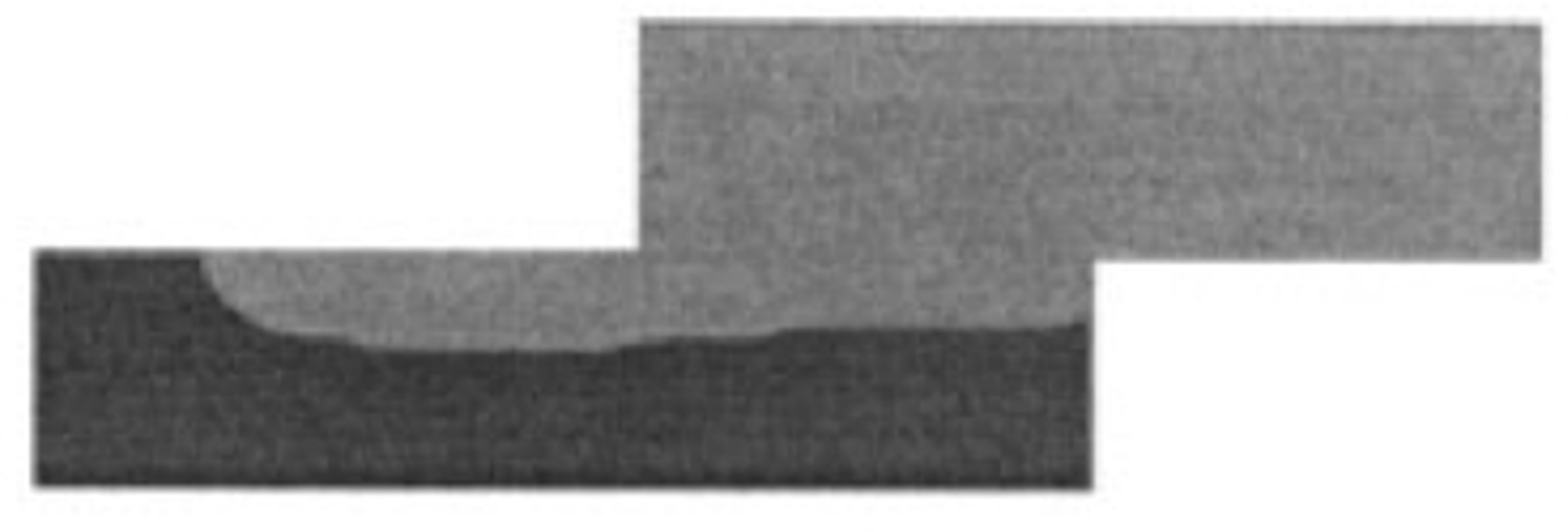}
        \label{fig:StepCavityRef2D08s}
    \end{subfigure}
    ~ 
    \begin{subfigure}[t]{0.3\textwidth}
        \includegraphics[width=\textwidth]{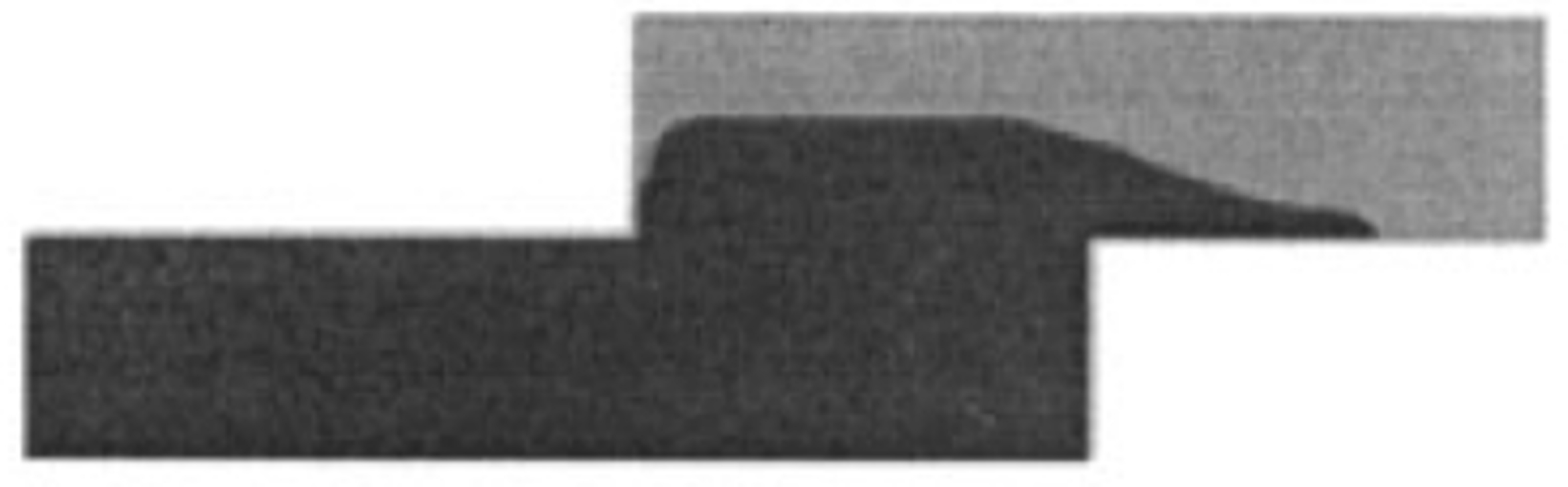}
        \label{fig:StepCavityRef2D16s}
    \end{subfigure}\\
        \begin{subfigure}[t]{0.3\textwidth}
        \includegraphics[width=\textwidth]{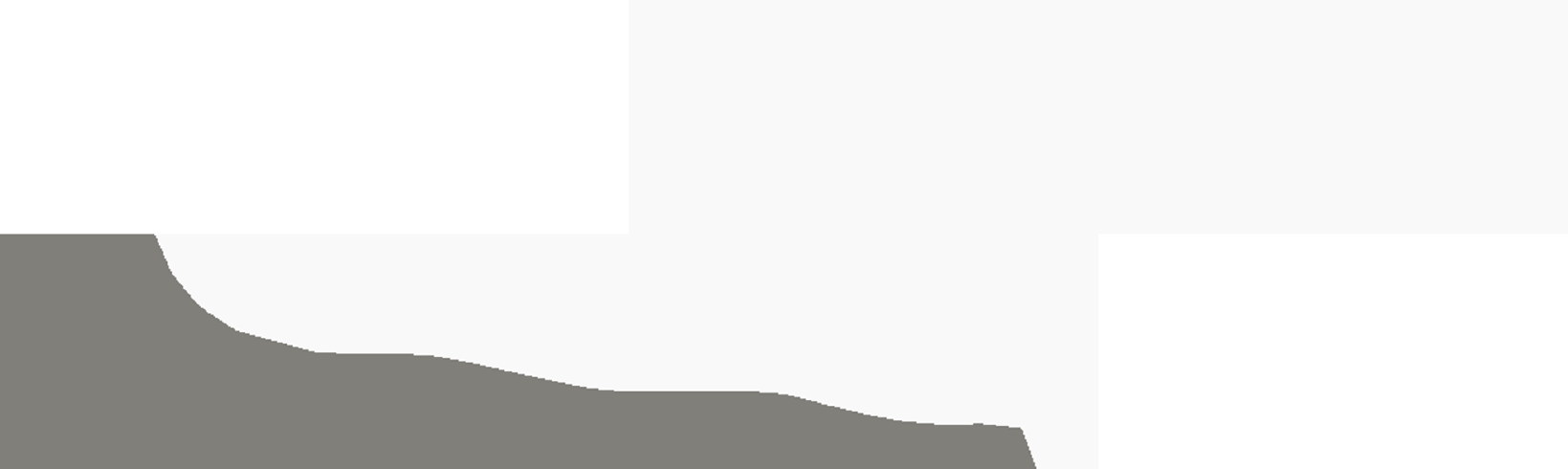}
        \caption{\(t= \SI{0.4}{s}\)}
        \label{fig:StepCavityRef2D04s}
    \end{subfigure}
    ~ 
    \begin{subfigure}[t]{0.3\textwidth}
        \includegraphics[width=\textwidth]{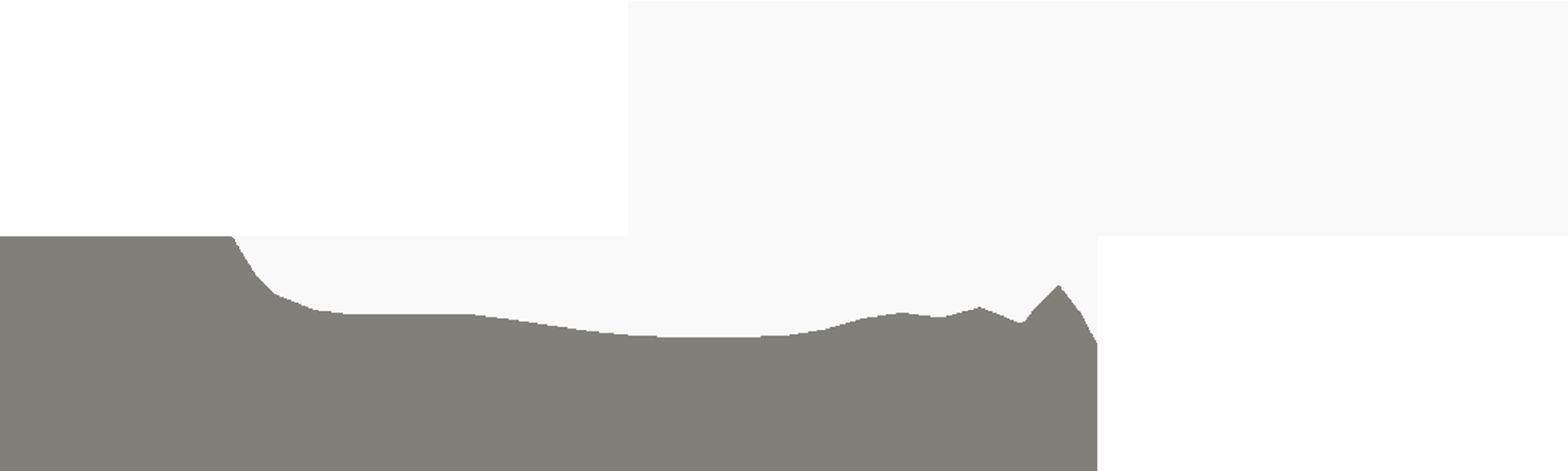}
        \caption{\(t= \SI{0.8}{s}\)}
        \label{fig:StepCavityRef2D08s}
    \end{subfigure}
    ~ 
    \begin{subfigure}[t]{0.3\textwidth}
        \includegraphics[width=\textwidth]{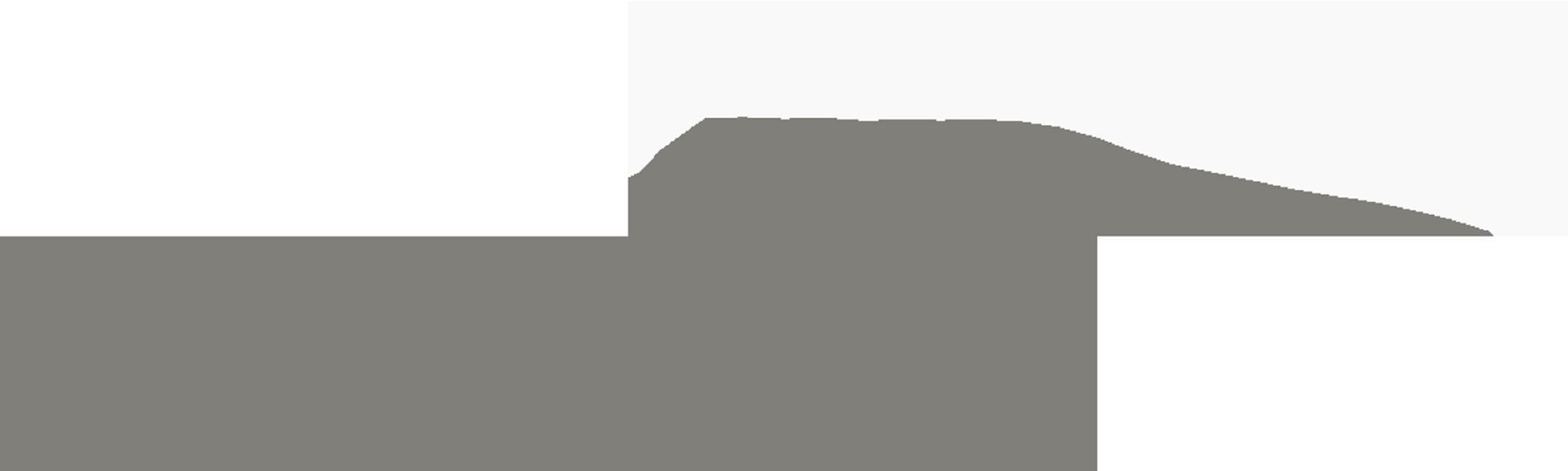}
        \caption{\(t= \SI{1.6}{s}\)}
        \label{fig:StepCavityRef2D16s}
    \end{subfigure}
    \caption{Molten material position at various time instances, obtained with a prism-type space-time discretization (top row) and a simplex-type space-time discretization (bottom row) and compared with reference data (middle row), published by \citet{Tezduyar2002}.}
    \label{fig:StepCavity2DShape}
\end{figure}

The molten material filled the mold first for \(t = \SI{1.6}{s}\) using the usual discontinuous-in-time Galerkin time stepping (prismatic space-time elements). These standard results were then compared with the filling results obtained with a hybrid tetrahedral-based space-time mesh discretization of the slab. This hybrid mesh was generated using the technique presented in \cite{behr2008simplex}, with each of the 320 time slabs being 0.005 thick and discretized differently with one to five elements in the time direction (cf. Figure \ref{fig:SpaceTimeSlabStepCavity}). The part of the domain, where temporal accuracy is increased, covers the area close to the propagating interface, as shown in Figure \ref{fig:SpaceTimeSlabStepCavity}.

Figure \ref{fig:StepCavity2DShape} illustrates the front position of the molten material at various time instances. The results obtained with the usual prismatic space-time discretization and with the hybrid mesh are also compared with those reported in Reference \cite{Tezduyar2002}. As we can see from Figure \ref{fig:StepCavity2DShape}, the results show a good agreement with the reference data \cite{Tezduyar2002}.
\section{Efficiency Aspects}
The example in Section \ref{StepCavity} is used to provide reliable timing measurements, although it can be considered small and not too complex. Table \ref{table:Performance} summarizes the typical performance behavior. The number of time steps, the number of nodes, the number of elements, as well as the total time required to form and to solve the equation systems are listed in Table \ref{table:Performance}, for 
\begin{enumerate*}[label={(\alph*)}]
\item a prism-based space-time finite element formulation with small time step (\(\Delta t = \SI{0.001}{s}\)),
\item a prism-based space-time finite element formulation with large time step (\(\Delta t = \SI{0.005}{s}\)), as used in Figure \ref{fig:StepCavity2DShape} (top row), and
\item a simplex-based space-time finite element formulation with variable time step (\(\Delta t = \SI{0.001}{}-\SI{0.005}{s}\)), as used in Figure \ref{fig:StepCavity2DShape} (bottom row).
\end{enumerate*}
The results of the prism-based space-time finite element formulation with small time step (\(\Delta t = \SI{0.001}{s}\)) are not presented in Subsection \ref{StepCavity}, but look similar to the reference data, shown in Figure \ref{fig:StepCavity2DShape} (middle row).
\begin{table}[ht]
\caption{Typical performance behavior of the prism- and simplex-based calculations.}
\centering
\resizebox{\textwidth}{!}{
\begin{tabular}{c c c c c c}
\toprule
 & Time  & Nodes    & Elements & System                 & System                \\
 & Steps & per step & per step  & formation (\si{s})  & Solution (\si{s})   \\ [0.5ex] 
\midrule
Prism,    \(\Delta t = \SI{0.001}{s}\)                     & 1600 &   730  &    624  & 331.65 & 5317.04\\
Prism,    \(\Delta t = \SI{0.005}{s}\)                     &   320 &   730  &    624  &   77.77 & 1260.55 \\
Simplex, \(\Delta t = \SI{0.001}{}-\SI{0.005}{s}\) &   320 & \(\SI{\sim 919}{}\)  & \(\SI{\sim  3003}{}\) & 142.21 & 1874.36 \\ [1ex]
\bottomrule
\end{tabular}}
\label{table:Performance}
\end{table}

According to the above table, the total time required to obtain a solution with time step \(\Delta t = \SI{0.001}{s}\) is four times that required to obtain a solution with time step \(\Delta t = \SI{0.005}{s}\), assuming fixed linear solver parameters and a prism-based space-time finite element formulation.  The average number of nodes in the hybrid mesh with variable temporal refinement is ca. \(919\), whereas the maximum and the minimum number of nodes is \(730\) and \(998\), respectively. That means that the solution time is also higher in the case of a simplex-based space-time finite element formulation with variable temporal refinement than the one required for a prismatic mesh with the same maximum time step size when using iterative solvers with linear scaling properties. It should be mentioned that at every time step, Newton-Raphson iterations are performed to solve the nonlinear discretized system resulting from the nonlinear Navier-Stokes equations and strong coupling iterations are also executed because of the mutual dependence of the level set field on the fluid velocity field and vice versa. A GMRES solver is used to solve the resulting linear system of equations. However, the conditioning of the linear systems arising from both types of elements was not examined in detail. The convergence of the iterative GMRES solver was similar in all the aforementioned test cases, though.

Furthermore, as the average number of elements in a simplex-based mesh is significantly higher than that in the prism-based meshes (ca. \(3003\) on average versus \(730\)), even though the elements are simpler, the system formation time takes almost twice as long. This time remains lower, however, than the time required to repeatedly form the system for a prism-based space-time finite element formulation with the small time step (\(\Delta t = \SI{0.001}{s}\)). 

To sum up, the use of simplex space-time meshes without any temporal refinement has some disadvantages. The equation system size remains the same in comparison with a prismatic mesh of the same time step size, but the number of elements is significantly increased leading to increased system formation times. However, the efficiency of the discretization is much improved when using local temporal refinement, while providing a better resolution of the space-time evolving interface. 

\section{Concluding remarks}
\label{ConcludingRemarks}

We have presented an updated version of the straightforward method for generating simplex space-time meshes, which was already introduced by \citet{behr2008simplex}. This version is based on the level-set method and allows arbitrary temporal refinement of the space-time slabs in the vicinity of evolving fronts. We have tested the resulting unstructured space-time meshes in the context of two-phase flow problems. The benchmark cases of the static bubble and the rising bubble/droplet have served as the initial validation of the unstructured space-time mesh solver for the Navier-Stokes equations and the level-set equation in two and three space dimensions. The benchmark case of a two-dimensional step cavity filling was used for checking the reliability of the results obtained with a hybrid mesh. Future work includes the extension of the arbitrary temporal refinement to more complicated problems, such as the filling of three-dimensional complex molds. The numerical behavior of stabilized FE formulations on such highly unstructured space-time meshes still needs to be examined. Likewise, the efficiency aspects are yet to be analyzed for more complex geometries.

\section*{Acknowledgements}
\label{Acknowledgements}

The authors gratefully acknowledge the support of the German Research Foundation (DFG) under program SFB 1120. The computations were conducted on computing clusters provided by the RWTH Aachen University IT Center and by the J\"ulich Aachen Research Alliance (JARA).


\begin{thebibliography}{24}

\bibitem[Adelsberger et~al.(2014)Adelsberger, Esser, Griebel, Gro{\ss}, Klitz,
  and R{\"u}ttgers]{adelsberger20143d}
J.~Adelsberger, P.~Esser, M.~Griebel, S.~Gro{\ss}, M.~Klitz,
  and A.~R{\"u}ttgers.
\newblock {3D} incompressible two-phase flow benchmark computations for rising
  droplets.
\newblock In \emph{Proceedings of the 11th World Congress on Computational
  Mechanics (WCCM XI), Barcelona, Spain, 2014.}, 2014.

\bibitem[Behr and Tezduyar(1994)]{behr1994finite}
M.~Behr and T.~E.~Tezduyar.
\newblock Finite element solution strategies for large-scale flow simulations.
\newblock \emph{Computer Methods in Applied Mechanics and Engineering},
  112\penalty0 (1):\penalty0 3--24, 1994.

\bibitem[Behr(2008)]{behr2008simplex}
M.~Behr.
\newblock Simplex space--time meshes in finite element simulations.
\newblock \emph{International journal for numerical methods in fluids},
  57\penalty0 (9):\penalty0 1421--1434, 2008.

\bibitem[Cruchaga et~al.(2002)Cruchaga, Celentano, and Tezduyar]{Tezduyar2002}
M.~Cruchaga, D.~Celentano, and T.~Tezduyar.
\newblock Computation of mould filling processes with a moving {L}agrangian
  interface technique.
\newblock \emph{Communications in Numerical Methods in Engineering},
  18\penalty0 (7):\penalty0 483--493, 2002.

\bibitem[Dhatt et~al.(1990)Dhatt, Gao, and Cheikh]{dhatt1990finite}
G.~Dhatt, D.~Gao, and A.~B.~Cheikh.
\newblock A finite element simulation of metal flow in moulds.
\newblock \emph{International Journal for Numerical Methods in Engineering},
  30\penalty0 (4):\penalty0 821--831, 1990.

\bibitem[Donea and Huerta(2003)]{donea2003finite}
J.~Donea and A.~Huerta.
\newblock \emph{Finite element methods for flow problems}.
\newblock John Wiley \& Sons, 2003.

\bibitem[Erickson et~al.(2005)Erickson, Guoy, Sullivan, and
  {\"U}ng{\"o}r]{erickson2005building}
J.~Erickson, D.~Guoy, J.~M.~Sullivan, and A.~{\"U}ng{\"o}r.
\newblock Building spacetime meshes over arbitrary spatial domains.
\newblock \emph{Engineering with Computers}, 20\penalty0 (4):\penalty0
  342--353, 2005.

\bibitem[Freudenthal(1942)]{freudenthal1942simplizialzerlegungen}
H.~Freudenthal.
\newblock Simplizialzerlegungen von beschrankter {F}lachheit.
\newblock \emph{Annals of Mathematics}, 43\penalty0 (3):\penalty0 580--582,
  1942.

\bibitem[Hansbo(1992)]{hansbo1992characteristic}
P.~Hansbo.
\newblock The characteristic streamline diffusion method for the time-dependent
  incompressible {N}avier-{S}tokes equations.
\newblock \emph{Computer Methods in Applied Mechanics and Engineering},
  99\penalty0 (2-3):\penalty0 171--186, 1992.

\bibitem[Hansbo and Szepessy(1990)]{hansbo1990velocity}
P.~Hansbo and A.~Szepessy.
\newblock A velocity-pressure streamline diffusion finite element method for
  the incompressible {N}avier-{S}tokes equations.
\newblock \emph{Computer Methods in Applied Mechanics and Engineering},
  84\penalty0 (2):\penalty0 175--192, 1990.

\bibitem[Hughes and Hulbert(1988)]{hughes1988space}
T.~J.~Hughes and G.~M.~Hulbert.
\newblock Space-time finite element methods for elastodynamics: formulations
  and error estimates.
\newblock \emph{Computer methods in applied mechanics and engineering},
  66\penalty0 (3):\penalty0 339--363, 1988.

\bibitem[Hughes et~al.(1987)Hughes, Franca, and Mallet]{hughes1987new}
T.~J.~Hughes, L.~P.~Franca, and M.~Mallet.
\newblock A new finite element formulation for computational fluid dynamics:
  {VI}. {C}onvergence analysis of the generalized supg formulation for linear
  time-dependent multidimensional advective-diffusive systems.
\newblock \emph{Computer Methods in Applied Mechanics and Engineering},
  63\penalty0 (1):\penalty0 97--112, 1987.

\bibitem[Hughes et~al.(1989)Hughes, Franca, and Hulbert]{hughes1989new}
T.~J.~Hughes, L.~P.~Franca, and G.~M.~Hulbert.
\newblock A new finite element formulation for computational fluid dynamics:
  {VIII}. {T}he {G}alerkin/least-squares method for advective-diffusive
  equations.
\newblock \emph{Computer Methods in Applied Mechanics and Engineering},
  73\penalty0 (2):\penalty0 173--189, 1989.

\bibitem[Hysing(2006)]{hysing2006new}
S.~Hysing.
\newblock A new implicit surface tension implementation for interfacial flows.
\newblock \emph{International Journal for Numerical Methods in Fluids},
  51\penalty0 (6):\penalty0 659--672, 2006.

\bibitem[Hysing et~al.(2009)Hysing, Turek, Kuzmin, Parolini, Burman, Ganesan,
  and Tobiska]{hysing2009quantitative}
S.-R.~Hysing, S.~Turek, D.~Kuzmin, N.~Parolini, E.~Burman,
  S.~Ganesan, and L.~Tobiska.
\newblock Quantitative benchmark computations of two-dimensional bubble
  dynamics.
\newblock \emph{International Journal for Numerical Methods in Fluids},
  60\penalty0 (11):\penalty0 1259--1288, 2009.

\bibitem[Lehrenfeld(2015)]{lehrenfeld2015nitsche}
C.~Lehrenfeld.
\newblock The nitsche {XFEM}-{DG} space-time method and its implementation in
  three space dimensions.
\newblock \emph{SIAM Journal on Scientific Computing}, 37\penalty0
  (1):\penalty0 A245--A270, 2015.

\bibitem[Neum{\"u}ller and Steinbach(2011)]{neumuller2011refinement}
M.~Neum{\"u}ller and O.~Steinbach.
\newblock Refinement of flexible space--time finite element meshes and
  discontinuous {G}alerkin methods.
\newblock \emph{Computing and visualization in science}, 14\penalty0
  (5):\penalty0 189--205, 2011.

\bibitem[Osher and Sethian(1988)]{osher1988fronts}
S.~Osher and J.~A.~Sethian.
\newblock Fronts propagating with curvature-dependent speed: {A}lgorithms based
  on {H}amilton-{J}acobi formulations.
\newblock \emph{Journal of computational physics}, 79\penalty0 (1):\penalty0
  12--49, 1988.

\bibitem[Sauerland and Fries(2013)]{sauerland2013stable}
H.~Sauerland and T.-P.~Fries.
\newblock The stable {XFEM} for two-phase flows.
\newblock \emph{Computers \& Fluids}, 87:\penalty0 41--49, 2013.

\bibitem[Shakib(1989)]{shakib1989finite}
F.~Shakib.
\newblock Finite element analysis of the compressible {E}uler and
  {N}avier-{S}tokes equations.
\newblock 1989.

\bibitem[Tezduyar et~al.(1992{\natexlab{a}})Tezduyar, Behr, Mittal, and
  Liou]{tezduyar1992newB}
T.~Tezduyar, M.~Behr, S.~Mittal, and J.~Liou.
\newblock A new strategy for finite element computations involving moving
  boundaries and interfaces--the deforming-spatial-domain/space-time procedure:
  {II}. {C}omputation of free-surface flows, two-liquid flows, and flows with
  drifting cylinders.
\newblock \emph{Computer Methods in Applied Mechanics and Engineering},
  94\penalty0 (3):\penalty0 353--371, 1992{\natexlab{a}}.

\bibitem[Tezduyar et~al.(1992{\natexlab{b}})Tezduyar, Behr, and
  Liou]{tezduyar1992newA}
T.~Tezduyar, M.~Behr, and J.~Liou.
\newblock A new strategy for finite element computations involving moving
  boundaries and interfaces--the deforming-spatial-domain/space-time procedure:
  {I}. {T}he concept and the preliminary numerical tests.
\newblock \emph{Computer methods in applied mechanics and engineering},
  94\penalty0 (3):\penalty0 339--351, 1992{\natexlab{b}}.

\bibitem[{\"U}ng{\"o}r and Sheffer(2000)]{ungor2000tent}
A.~{\"U}ng{\"o}r and A.~Sheffer.
\newblock Tent-pitcher: A meshing algorithm for space-time discontinuous
  {G}alerkin methods.
\newblock In \emph{{I}n {P}roc. of 9th {I}nt\'{}l. {M}eshing {R}oundtable},
  2000.

\bibitem[Wang and Persson(2013)]{wang2013discontinuous}
L.~Wang and P.-O.~Persson.
\newblock A discontinuous {G}alerkin method for the {N}avier-{S}tokes equations
  on deforming domains using unstructured moving space-time meshes.
\newblock \emph{AIAA Paper}, 2833:\penalty0 2013, 2013.

\end{thebibliography}
\end{document}